\documentclass[a4paper,12pt]{article}

\usepackage{enumitem} 
\usepackage[utf8]{inputenc} 
\usepackage{amsmath, amsthm, amssymb, amsfonts,mathtools,physics,bbm}
\numberwithin{equation}{section}
\usepackage{subcaption}
\usepackage[multiple]{footmisc}
\usepackage{float}
\usepackage[textsize=scriptsize]{todonotes}
\usepackage[hmargin=1in,vmargin=1.25in,bindingoffset=0in]{geometry}

\usepackage[numbers,sort&compress]{natbib}
\usepackage[linktoc=page,colorlinks]{hyperref}
\hypersetup{
    colorlinks=true,
    urlcolor=black, 
    citecolor=teal
}
\allowdisplaybreaks

\usepackage{tikz}

\def\d{\partial}

\def\cp{\mathbb{CP}}
\def\notwo{\cN=(0,2)}
\def\bphi{\overline{\varphi}}
\def\tphi{\tilde{\varphi}}
\def\bpsi{\overline{\psi}}

\def\arc{\operatorname{arc}\hspace{-2pt}}

\def\srb{S[\varphi_{\rm rb}]}
\def\srbz{S_{\rm rb}^{(0)}}

\renewcommand*{\tilde}[1]{\widetilde{#1}}
\def\aB#1{\alpha_{B #1}}
\def\tB#1{\tau_{B #1}}
\def\ms{m_{*}}

\DeclareMathOperator{\cn}{cn}
\DeclareMathOperator{\sn}{sn}
\DeclareMathOperator{\dn}{dn}
\DeclareMathOperator{\cs}{cs}
\DeclareMathOperator{\cd}{cd}
\DeclareMathOperator{\sd}{sd}

\def\oFt#1{{}_{1}F_{2}\left(\textstyle #1 \right)}
\def\GB#1{G_B\left( #1 \right)}

\def\RR{\mathbb{R}}

\def\ZZ{\mathbb{Z}}

\def\cB{\mathcal{B}}
\def\cC{\mathcal{C}}
\def\cD{\mathcal{D}}

\def\cG{\mathcal{G}}

\def\cI{\mathcal{I}}
\def\cJ{\mathcal{J}}
\def\cK{\mathcal{K}}
\def\cL{\mathcal{L}}

\def\cN{\mathcal{N}}

\def\cR{\mathcal{R}}

\begin{document}

\begin{titlepage}

\begin{flushright}
UCI-TR-2024-18
\end{flushright}
    
\vspace{4mm}
    
\begin{center}
{\Large \bf  
Nonperturbative features in the
\\[4mm]
Lie-algebraic K\"ahler sigma model with fermions
}

\vspace{6mm}
    
{\large \bf   
Chao-Hsiang Sheu$^{a}$ 
}
\end {center}
    
\begin{center}
    
{\it  
$^{a}$Department of Physics and Astronomy, University of California,\\
Irvine, CA 92697-4575, U.S.A.
}
\end{center}
    
\vspace{1cm}
    
\begin{center}
{\large\bf Abstract}
\end{center}
We investigate the trans-series structure of a quantum mechanical system originating from a Lie-algebraic K\"ahler sigma model with multiple right-handed chiral fermions, extending previous results for the standard onecomplex projective ($\cp^1$) model \cite{Fujimori:2016ljw, Fujimori:2017oab} to its deformed counterpart. We identify and analyze saddle point solutions and examine their contributions within the perturbative expansions of the ground state energy, revealing that the ambiguity structure observed in the $\cp^1$ model persists in the deformed model as well. Additionally, we explore the role of the elongation parameter and its potential impact on higher-loop corrections, and propose that it becomes relevant in shaping the system's quantum behavior from the three-loop level. This verifies that the trans-series framework provides a comprehensive approach to capturing the structure of quantum fluctuations and ambiguities in these deformed sigma models.

\end{titlepage}
\newpage

\tableofcontents

\section{Introduction}

Sigma models provide a powerful framework not only for modeling real-world physical phenomena but also as a versatile theoretical laboratory that bridges concepts in both physics and mathematics. In this work, we delve into a specific class of deformed sigma models preserving $U(1)$ symmetry, with a K\"ahlerian target space. These models extend the familiar $\cp^1$ geometry in a Lie-algebraic manner \cite{Gamayun:2023atu,Sheu:2023hoz,Gamayun:2023sif,SS2023}. Here, we will further investigate the quantum mechanical properties of the deformed $\cp^1$ model by compactifying it on a cylinder, which allows us to capture key quantum features and examine the semiclassical configurations of the deformed geometry in a lower-dimensional setup.

In the 1990s, studies on related types of deformations gained attention through researches on two-dimensional black hole solutions \cite{Witten:1991yr} and the integrable deformation of the $O(3)$ sigma model \cite{Fateev:1992tk}. In particular, the latter investigation was extended to include a broader class of one-parameter integrable deformations for principal chiral models \cite{Klimcik:2002zj, Klimcik:2008eq} and coset models \cite{Sfetsos:2015nya, Delduc:2013fga, Bykov:2020llx}. These deformations are collectively referred to as Yang-Baxter models or $\eta$-deformations, distinguished by their preservation of integrable structure even under deformation. In recent years, these models have attracted renewed interest, largely due to their relevance in the context of holographic duality, particularly in the AdS/CFT correspondence \cite{Delduc:2014kha, PhysRevLett.112.051601}. A comprehensive review of developments in this field can be found in \cite{Hoare:2021dix}. The classical integrability properties of sigma models have since been extended to quantum integrability \cite{Bazhanov:2018xzh, Bazhanov:2017nzh}, allowing a deeper exploration of the interplay between deformation parameters and algebraic structures. Notably, by introducing an elongation parameter $k$ in the Lie-algebraic parametrization, with $k \in [1, \infty)$ defined as $k=(1-\eta^2)/(1+\eta^2)$, it becomes possible to show that the deformed $\cp^1$ model aligns with the $\eta$-deformed version of the $\cp^1$ model. This connection reveals interesting structural features that make the deformed models particularly suitable for exploring solvability of the model, especially when starting with simpler quantum mechanical systems as a foundational approach.

To analyze quantum systems beyond integrable approaches, many insights can be gained by examining different (asymptotic) perturbative trans-series \cite{Ecalle} of relevant physical quantities. In recent years, this method has been intensively developed within the framework of resurgence theory, with applications across diverse areas in physics. For instance, resurgence techniques have been successfully applied to quantum mechanical systems \cite{Basar:2015xna, Behtash:2015loa,Fujimori:2016ljw,Fujimori:2017oab,Fujimori:2017osz, Fujimori:2022lng}, field theories both with and without integrable structures \cite{Dunne:2012ae,Cherman:2013yfa,Dunne:2015ywa,Demulder:2016mja,Fujimori:2018kqp,Schepers:2020ehn,Marino:2021six,DiPietro:2021yxb,Marino:2021dzn,Marino:2022ykm}, and even to topological string theory \cite{Couso-Santamaria:2014iia,Gu:2021ize,Gu:2022sqc,Gu:2023mgf,Marino:2023gxy,Marino:2024yme}.
This resurgence framework is further applicable in the problems to be considered based on a recent observation. As shown in \cite{SS2023}, the Lie-algebraic $\cp^1$ model can be continuously mapped to the Lam\'e system \cite{Lame,Dunne:2002at,NIST:DLMF}, which was previously solved in \cite{Basar:2015xna}, through a specific compactification scheme \cite{Scherk:1979zr}. This connection highlights new potential pathways for solving the deformed $\cp^1$ quantum mechanics within the Kaluza-Klein (KK) compactification framework. Importantly, the Lam\'e system is not a simple zero angular momentum limit of the deformed $\cp^1$ model, as it includes a nontrivial linear differential term \cite{SS2023}. This distinction underscores the rich structure that emerges when exploring these systems via trans-series techniques.

This paper aims to investigate the Lie-algebraic sigma model from a quantum mechanical point of view, building on the approach outlined in \cite{Fujimori:2016ljw, Fujimori:2017oab}. Previous research has extensively examined principal chiral models \cite{Cherman:2013yfa, Dunne:2015ywa} and their deformations \cite{Demulder:2016mja, Schepers:2020ehn}, as well as the coset models \cite{Dunne:2012ae, Fujimori:2016ljw, Fujimori:2017oab, Fujimori:2017osz}. In this work, we focus specifically on deformations of the $\cp^1$ model, incorporating multiple fermions. We analyze ground state energy perturbations around the supersymmetric point, employing the path integral formalism to capture quantum fluctuations accurately.
Two perspectives, based on weak coupling and near-supersymmetry limits, converge to reveal a comparable structure of ambiguities as seen in the standard $\cp^1$ model. Additionally, we conjecture that the elongation parameter $k$ starts exerting a significant influence at the three-loop level, indicated by the term $g^4(k^2-1)$. This suggests that higher-order quantum corrections may depend intricately on $k$, potentially unveiling new structural features within the deformed model.

The paper is organized as follows. In Sec. \ref{sec:model}, we set up the quantum mechanical model derived from the $\cN=(0,2)$ Lie-algebraic sigma model to study in this paper. The saddle point solutions, including real and complex bions are given. In Sec. \ref{sec:epi}, we calculate the corrections to the ground state energy arising from quantum fluctuations using the path integral formulation, and we identify the contributions from each saddle point. Next, the ground state energy near the supersymmetric vacuum is investigated by the perturbation expansion in Sec. \ref{sec:nsr} and compared with the result obtained previously. Conclusions and some further discussions are given in Sec. \ref{sec:clsn}. Conventions and supplementary materials are provided in the appendices.

\section{Defining the model}
\label{sec:model}

In this section, we present the quantum mechanical model that forms the basis of our analysis, establishing its connection to the two-dimensional Lie-algebraic sigma model. Additionally, we provide saddle point solutions to support the discussions that follow.

\subsection{The deformed $\cp^1$ model}

In the first part, we review the geometric properties of our main model previously constructed in \cite{Sheu:2023hoz,Gamayun:2023sif,Gamayun:2023atu} and derive its extension with left-handed fermions.

The deformed $\cp^1$ model arises from the Lie-algebraic metric of $sl(2) \times sl(2)$ algebra with its non-linear realization, say,
\begin{align}
    T^+ = -\varphi^2\pdv{\varphi} \,,\quad 
    T^0 = \varphi\pdv{\varphi} \,,\quad
    T^- = \pdv{\varphi}
\end{align}
and their complex conjugates. $T^{\pm}$ and $T^0$ are the generators of $sl(2)$ algebra. Assuming the $U(1)$ symmetry of the system is preserved, we can construct a generic metric corresponding to a linear combination of quadratic operators
\begin{align}
    G^{1\bar{1}} \pdv{\varphi}\pdv{\bphi{}} 
    = n_1 T^-\overline{T}^- + n_2 T^0\overline{T}^0 + n_3 T^+\overline{T}^+ \,.
\end{align}
In addition, as long as the parameters $n_{1,3}$ are non-singular (i.e. neither 0 nor $\infty$), the metric $G$ can be simplified by rescaling $\varphi$ and $\bphi{}$. Namely,
\begin{align}
    G \equiv G_{1\bar{1}} = \frac{2}{g^2}\frac{1}{1 + 2k \abs{\varphi}^2 + \abs{\varphi}^4}
\end{align}
i.e. 
\begin{align}
    n_1 = n_3 = \frac{g^2}{2} 
    \qand
    n_2 = g^2k
\end{align}
with $k$ being the elongation constant. 

\begin{figure}[t] 
  \centering 
  \includegraphics[width=.5\linewidth]{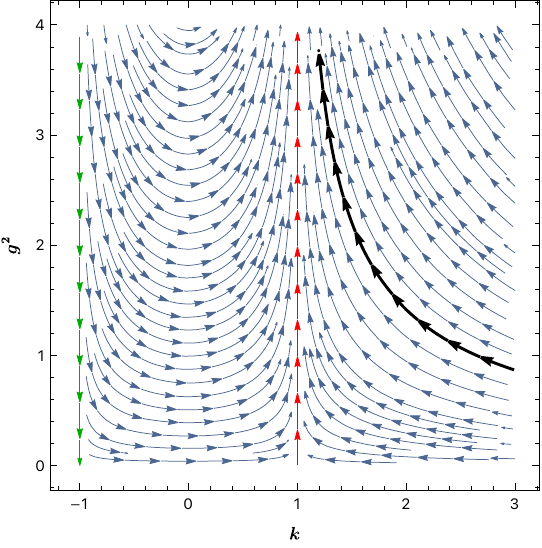}
  \caption{\small The RG flow of $g^2$ and $k$ couplings. The flow is from UV to IR. The separatrices are marked with green and red lines while the thick black line represents an example of the RG flow in the limit of the  sausage model  (i.e. large $k$). See for example \cite{Fateev:1992tk,Lukyanov:2019asr} for the illustration of the sausage model.}
  \label{fig:rgflow}
\end{figure}

Including the quantum corrections, the Lie algebraic sigma model is renormalizable \cite{Sheu:2023hoz,Gamayun:2023atu,Gamayun:2023sif} in two dimensions with the (one-loop) renormalization group (RG) flow equations in $g^2$ and $k$
\begin{align}
    \dv{g^2}{\log\mu} = -\frac{kg^4}{2\pi} \,,\quad
    \dv{(g^2k)}{\log\mu} = -\frac{g^4}{2\pi}\,.
\end{align}
Two coupling constants are entangled and the combination $g^4(k^2-1)$ is a RG-invariant.
Figure \ref{fig:rgflow} illustrates the direction of the renormalization group (RG) flow from the ultraviolet (UV) scale to the infrared (IR) scale. The separatrices are located at $k=\pm 1$, and in particular, the streamline at $k=1$ corresponds to the standard $\cp^1$ model.
In the $k \geq 1$ sector, the initial state with a small $g^2$ coupling follows the RG flow to a state with strong coupling. On the contrary, when the RG flow starts between the separatrices ($-1\leq k \leq 1$), the coupling initially weakens but eventually becomes strong as the elongation parameter approaches unity. The theory is asymptotically free and restore the full $O(3)$ symmetry in the IR regime.

\subsubsection*{Adding fermions}

To get fermions involved in the model, let us start with considering its $\cN=(0,2)$ supersymmetrization since we would like to consider the perturbation theory around this supersymmetric point in the subsequent discussion. Along the same line with \cite{Witten:2005px,Shifman:2008wv,Chen:2014efa}, we know that the Lagrangian of a minimal $\notwo$ sigma model takes the form
\begin{align}\label{eq:2dLag}
    \cL_{\rm 2d} 
    = \frac{G}{g^2_{\rm 2d}} \biggl[
        \d^\mu\varphi\d_{\mu}\bphi{}
        + \bpsi{}i\d_{R}\psi + i\bpsi{}\Gamma\psi(\d_R\varphi)
    \biggr]  
\end{align}
in which $\psi$ is a left-handed fermion and $\d_R$ is the right-handed coordinate differentiation. The Christoffel symbol is given by the following expression
\begin{align}
    \Gamma  = \pdv{}{\varphi}\log{G}
    = -\frac{2\bphi{}(k+\abs{\varphi}^2)}{1+2k\abs{\varphi}^2+\abs{\varphi}^4}
    \,.
\end{align}
Then, we extend the \eqref{eq:2dLag} beyond the supersymmetric point by incorporating $N_f-1$ additional flavors\footnote{The notion of flavor in the present context differs from that in \cite{Cui:2011uw}, in which $N_f$ flavors of \emph{left-handed} fermions are introduced in a manner preserving the $(0,2)$ supersymmetry of the theory.} of right-handed fermions. Namely, 
\begin{align}
    \bpsi{}i\d_{R}\psi + i\bpsi{}\Gamma\psi(\d_R\varphi)
    \to 
    \sum_{a=1}^{N_f}\, \bpsi{}_ai\d_{R}\psi_a + i\bpsi{}_a\Gamma\psi_a(\d_R\varphi)
    \,.
\end{align}

In this scheme, the two dimensional theory is modified by the quantum effects and the beta function with $N_f$ Weyl fermions up to two loops \cite{SS2023} is 
\begin{align*}
    \beta(G) = \frac{1}{4\pi}G\cR + \frac{2-N_f}{32\pi^2}G\cR^2 + \cdots
\end{align*}
where $\cR$ stands for the scalar curvature of the target manifold.
In particular, the second order beta function vanishes as $N_f$ is set to 2. Indeed, the theory enjoys a $\cN=2$ supersymmetry for which the beta function is exhausted at one loop when the interaction $R_{1\bar{1}1\bar{1}}(\bpsi{}\psi)^2$ is included.\footnote{At the two-loop level, this vertex plays no roles in determining the beta function. An analogous example of the exact and perturbative beta functions of the $\cp^{N-1}$ model with its variation and homogenous spaces can be found in \cite{Cui:2010si,Cui:2011uw,Chen:2014efa,Sheu:2019eis}, respectively.}
On the other hand, for $N_f = 1$, the theory possesses the exact $\cN=(0,2)$ supersymmetry. 
This heterotic supersymmetric configuration serves as the foundation for the perturbative expansions developed in the upcoming sections. Specifically, we will incorporate interactions of the form $\bpsi{}_a\Gamma\psi{}_a\d_R\varphi$, representing the coupling between each fermion field $\psi_a$ and the bosonic field $\varphi$.

\subsection{Effective quantum mechanical model}

To derive the effective quantum mechanical description of the deformed model, we first take the theory on $\RR \times S^1_{\ell}$ into account via the KK decomposition\footnote{
If, instead, one considers the Scherk-Schwarz compactification scheme, the resulting effective quantum mechanics corresponds to the Lam\'e system, which has been shown to be Lie-algebraically solvable in both the bosonic \cite{Dunne:2002at} and supersymmetric \cite{SS2023} cases for specific parameter sets.} with the $\ZZ_2$ twisted-boundary condition \cite{Fujimori:2016ljw,Dunne:2012ae}. In choosing the boundary condition, one can explore the more detailed structure of the theory \cite{Dunne:2012ae,Cherman:2013yfa,Dunne:2015ywa,Demulder:2016mja,Fujimori:2016ljw,Fujimori:2017oab,Schepers:2020ehn}, particularly bions which will play crucial role in the discussion of the resurgence structure later on.
More specifically, the decomposition of the complex field and the fermion field is
\begin{align}\label{eq:sfbc}
    \varphi = \sum_{n \in \ZZ} \varphi_{(n)}(t)\exp[i\frac{2\pi}{\ell}\left( n+\frac{1}{2} \right)x] 
    \,,\qquad
    \psi = \sum_{n \in \ZZ} \psi_{(n)}(t)\exp[i\frac{2\pi}{\ell}\left( n+\frac{1}{2} \right)x]
\end{align}
where $\ell$ is the circumference of the compactified dimension.
Plugging \eqref{eq:sfbc} into \eqref{eq:2dLag} and integrating over the $x$-direction, one would find an equivalent theory with an infinite tower of Kaluza-Klein modes. Because we are interested in the semiclassical aspects, in the low energy regime, only the lowest mode $n=0$ is considered. Therefore, the Lagrangian of the one-dimensional model has the form 
\begin{align}\label{eq:qmpre}
    \cL = \frac{G}{g^2} \left[  
        \d_t\varphi\d_t\bphi{} - m^2\varphi\bphi{} + \sum_{a=1}^{N_f} i \bpsi{}_a\left( \d_t + \Gamma\d_t\varphi \right) \psi_a
        +  \sum_{a=1}^{N_f} m\left(  1+
            \frac{\varphi\Gamma + \bphi{}\Gamma^{\dagger}}{2}
        \right)\bpsi{}_a\psi_a
    \right]
\end{align}
where the parameters $g^2$ and $m$ are defined as
\begin{align}\label{eq:comppara}
    \frac{1}{g^2} = \frac{\ell}{g^2_{\rm 2d}}
    \,,\qquad
    m = \frac{\pi}{\ell}
    \,,
\end{align}
and $a=1,2,\cdots,N_f$ stands for the flavors of fermions.

In accordance with \cite{Behtash:2015loa}, the Hilbert space is graded by the fermion number in a quantum mechanical system with fermions. We can therefore project the original model onto its subspace with a fixed fermion number. It is then suffices to analyze solely the bosonic type theory with the potential derived from the fermions. The Hamiltonian subjected to the ground energy sector \cite{Fujimori:2017osz} takes the form 
\begin{align}\label{eq:hgrd}
    H= G^{-1}\left[ g^2 p_{\varphi}p_{\bphi{}} + \frac{1}{g^2}\abs{\pdv{\mu}{\varphi}}^2 - \epsilon\pdv{\mu}{\varphi}{\bphi{}}\right]
\end{align}
where $p_{\varphi}$ and $p_{\bphi{}}$ are the canonical momenta and $\mu$ is the moment map 
\begin{align}\label{eq:momentmap}
    \mu &= 
    \frac{m}{2\sqrt{k^2-1}} 
    \log(\frac{\abs{\varphi}^2+k-\sqrt{k^2-1}}{\abs{\varphi}^2+k+\sqrt{k^2-1}})
    + \frac{m\log(k+\sqrt{k^2-1})}{\sqrt{k^2-1}}
\end{align}
in which the constant of the moment map is chosen for convenience. 
Note that we replace the notation $N_f$, the number of flavors, with a continuous parameter $\epsilon$, regarding the term as a deformed term in the potential. The explicit expression of the potential term is 
\begin{align}\label{eq:potential}
    V(\varphi,\bphi{}) 
    = \frac{m^2}{g^2}\frac{\abs{\varphi}^2}{1+2k\abs{\varphi}^2+\abs{\varphi}^4}
    - \epsilon m \frac{1-\abs{\varphi}^4}{1+2k\abs{\varphi}^2+\abs{\varphi}^4}
    \,,
\end{align}
from which implies the time-independent Schr\"odinger equation
\begin{align}
    -G^{-1} \pdv[2]{\Psi}{\varphi}{\bphi{}} + (V(\varphi,\bphi{}) - E) \Psi = 0
\end{align}
In the limit $\epsilon \to 1$, the system is supersymmetric and hence the ground state energy vanishes exactly with the wave function 
\begin{align}
    \Psi^{(0)} = \exp(-\frac{\mu}{g^2})
\end{align}
i.e. $H\Psi^{(0)} = 0$ as expected. This assertion holds true for all models built from the chiral multiplets \cite{Fujimori:2017osz}.

Prior to deriving the saddle point solutions, we present a different parametrization of the deformed model embedded in a three-dimensional space. Namely,
\begin{align}\label{eq:ellpara}
    \varphi(\theta,\kappa) = 
    \frac{\sqrt{1-\kappa}\sd(\theta,\kappa)}{1+\cd(\theta,\kappa)}
    e^{i\alpha}
    \qand
    \bphi{}(\theta,\kappa) = 
    \frac{\sqrt{1-\kappa}\sd(\theta,\kappa)}{1+\cd(\theta,\kappa)}
    e^{-i\alpha}
\end{align}
where $\theta \in [0,2K(\kappa))$ is the angle from the $z$-axis and $\alpha \in [0,2\pi)$ is the azimuthal angle. $\cd$ and $\sd$ are Jacobi elliptic functions; see Appendix \ref{sec:nc} for their definitions. The elongation parameter $k$ is reparametrized as $\kappa$ in the way 
\begin{align}
    \kappa = \frac{k-1}{k+1} \,.
\end{align}
This can be thought of as an elliptic version of the standard stereographic projection. Unlike the traditional two-sphere target space of the $\cp^1$ model, the resulting geometry resembles a sausage shape \cite{Sheu:2023hoz,Fateev:1992tk}. 
Additionally, with \eqref{eq:ellpara}, the profile of the bion solutions can be visualized in a straightforward manner. Indeed, given a saddle point solution $\varphi_s(\tau)$, $\theta(\tau)$ turns out to be 
\begin{align}
    \theta(\tau) = \cd^{-1}\left( \frac{1-\varphi_s\bphi{}_s}{1+\varphi_s\bphi{}_s}, \kappa \right)
    \,.
\end{align}
It then suffices to consider the height function \cite{Fujimori:2016ljw}
\begin{align}\label{eq:heightf}
    \Sigma(\tau) = \frac{1-\varphi_s\bphi{}_s}{1+\varphi_s\bphi{}_s}
\end{align}
in the subsequent plots as $\Sigma(\tau)$ already summarizes all essential characteristics of the bion solutions.

\subsection{Saddle point solutions}

To advance our study of the corrections to the ground state energy due to the deformation of the boson-fermion interaction $\epsilon$ via the path integral, we would need the explicit constructions of the saddle point solutions.

Taking the Euclidean signature, we consider the action of the form 
\begin{align}\label{eq:actionqm}
    S = \int~
        \left( \frac{G}{g^2} \abs{\dot{\varphi}}^2
        + \frac{1}{g^2}\abs{\pdv{\mu}{\varphi}}^2 - \epsilon\pdv{\mu}{\varphi}{\bphi{}} \right)
    ~\dd{\tau}
\end{align}
where $\tau$ is the Euclidean time and $\mu$ is given in \eqref{eq:momentmap}. 
The equation of motion is then 
\begin{align}\label{eq:eom}
    &G \ddot{\bphi{}} + G^{(0,1)} \dot{\bphi{}}{}^2 -g^2\pdv{V}{\varphi} = 0
    \,,
\end{align}
and its complex conjugate.
It is worth to mention that the system is invariant under the time translation and the phase rotation, which leads to the conserved energy $E$ and angular momentum $J$, respectively. Namely,
\begin{align}\label{eq:EandJ}
    E = T^{00}
    = \frac{G}{g^2}\abs{\dot{\varphi}}^2 - V(\varphi) 
    \qand
    J = \frac{iG}{g^2}\left( \dot{\varphi}\bphi{} - \varphi\dot{\bphi{}} \right)
    \,.
\end{align}
With regard to the saddle point solutions, their actions are finite and subjected to the boundary condition
\begin{align}\label{eq:infbc}
    \lim_{\tau \to \pm\infty} \varphi 
    = \lim_{\tau \to \pm\infty} \bphi{}
    = 0
    \,.
\end{align}
Note that this implies that the saddle point solutions have zero angular momentum, $J=0$, and the potential's minimum occurs at $\tau \to \pm\infty$, as evident from Eq. \eqref{eq:potential}.
By combining the conserved charges and the boundary conditions, the equation of motion of the saddle point solutions can be reduced from \eqref{eq:eom} to its first integral 
\begin{align}\label{eq:saddle1}
    E = E(\tau \to \pm\infty) = \epsilon m \,.
\end{align}
Now, we can now proceed to demonstrate the following two kinds of solutions.

\subsubsection{Single bion configuration}

In the current model, there are two distinct bion solutions, the real bion and the complex bion. The latter one can be viewed as the complexification of the previous one in the parameter space.

\subsubsection*{Real bion solution}

Let us start with the real bion configuration. The solution to Eq. \eqref{eq:saddle1} takes the form 
\begin{align}\label{eq:realbion}
    \varphi_{\rm rb}(\tau) = e^{i\alpha}\sqrt{\frac{k\omega_k^2}{\omega_k^2-m^2}}\frac{1}{\sinh(\omega_k(\tau-\tau_0))}
\end{align}
where $\tau_0$ is the position of the real bion while $\alpha$ is its phase, showing the moduli space is a cylinder $\RR \times S^1$. The frequency $\omega_k$ is 
\begin{align}
    \omega_k = m \sqrt{1 + \frac{2k\epsilon g^2}{m}}
    \,.
\end{align}

\begin{figure}[t] 
    \centering 
    \begin{subfigure}[b]{.47\linewidth}
      \centering
      \includegraphics[width=.9\textwidth]{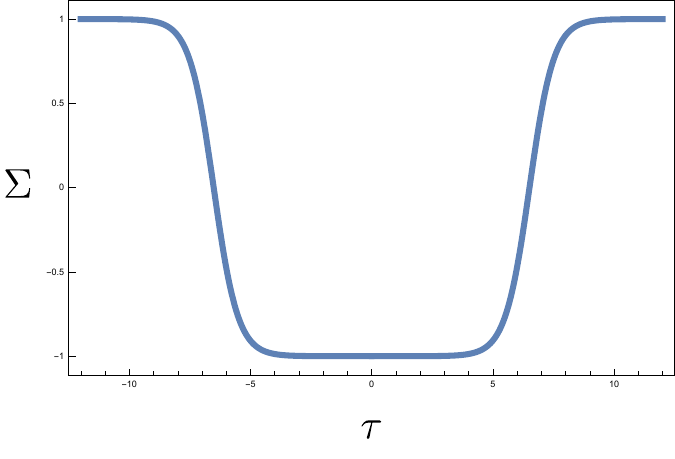}
      \caption{\small $k=2$}
    \end{subfigure}
    \begin{subfigure}[b]{.47\linewidth}
      \centering
      \includegraphics[width=.9\textwidth]{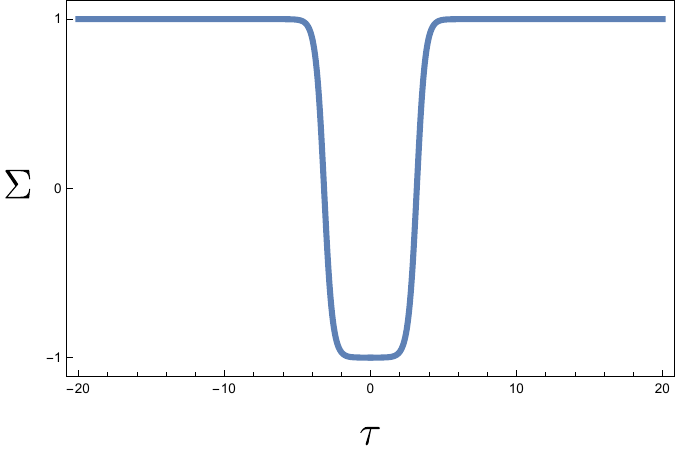}
      \caption{\small $k=10^6$}
    \end{subfigure}
    \caption{\small The profile of the real bion solution. The parameters are given as follows. $m=1, \epsilon=1.1, g=1/500$.}
    \label{fig:rbprofile}
\end{figure}
{\noindent}The profiles of $\varphi(\tau)$ are plotted in figure \ref{fig:rbprofile} in terms of the height function given in \eqref{eq:heightf}. It can be seen from the figure that the bion is composed of a kink and an anti-kink and their relative proximity gets closer as $k$ grows.
Indeed, the real bion solution \eqref{eq:realbion} can be recast as the kink-anti kink pair, namely,
\begin{align}\label{eq:rbkak}
    \varphi_{\rm rb} = \left[  
        e^{\omega_k(\tau - \tau_+) - i\alpha_+} + e^{-\omega_k(\tau - \tau_-) - i\alpha_-}
    \right]^{-1}
\end{align}
with its complex conjugate as the $\cp^1$ case \cite{Fujimori:2016ljw,Fujimori:2017oab,Fujimori:2017osz}.
Note that $\tau_{\pm}$ and $\alpha_{\pm}$ are the positions and phases of kink and anti kink, respectively. The parameters take the form 
\begin{align}\label{eq:rbr}
    \tau_{\pm} = \tau_0 \pm \frac{1}{2\omega_k}\log\frac{4k\omega_k^2}{\omega_k^2-m^2}
    \qand
    \alpha_{\pm} = \left( \alpha - \frac{\pi}{2} \right) \pm \frac{\pi}{2}
    \,.
\end{align}
This representation will be useful to clarify the meaning of the path integral realization of the perturbed ground state energy by the deformed potential shown in Sec. \ref{sec:nsr}.

\begin{figure}[t] 
    \centering 
    \begin{subfigure}[b]{.32\linewidth}
      \centering
      \includegraphics[width=.9\textwidth]{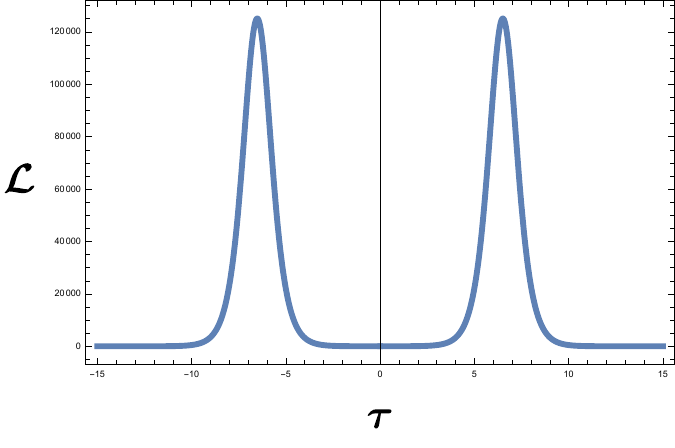}
      \caption{\small $k=1$}
    \end{subfigure}
    \begin{subfigure}[b]{.32\linewidth}
      \centering
      \includegraphics[width=.9\textwidth]{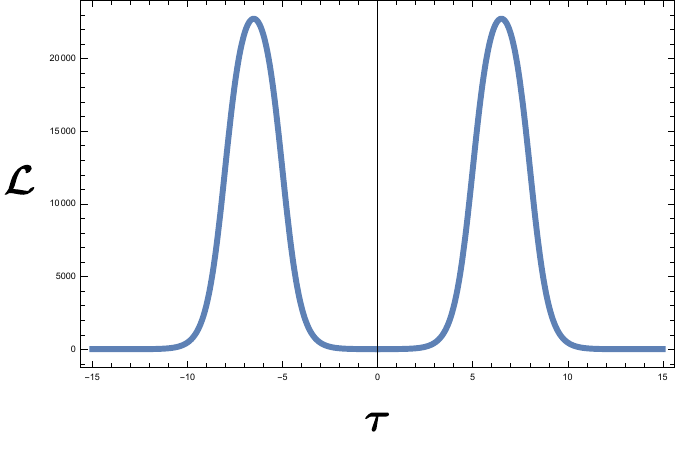}
      \caption{\small $k=10$}
    \end{subfigure}
    \begin{subfigure}[b]{.32\linewidth}
        \centering
        \includegraphics[width=.9\textwidth]{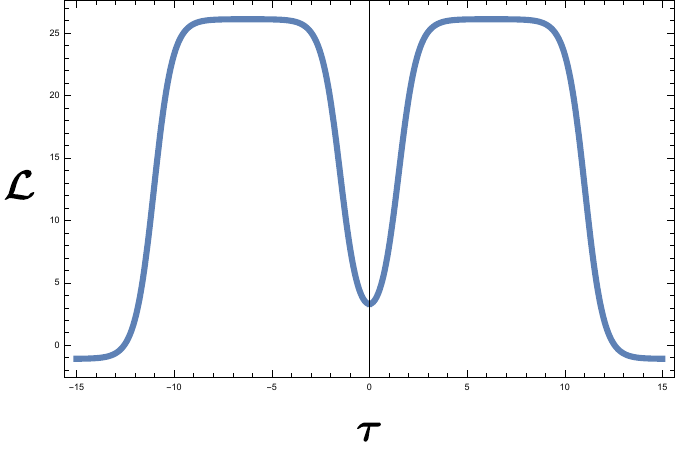}
        \caption{\small $k=10^4$}
      \end{subfigure}
    \caption{\small The Lagrangian over Euclidean time $\tau$. The parameters are given as follows. $m=1, \epsilon=1.1, g=1/500$.}
    \label{fig:lagrb}
\end{figure}

Then we can read off the Lagrangian of the real bion solutions
\begin{align}
    \cL_{\rm rb} = 
    4m\epsilon 
    \frac{4m\epsilon k^2\omega_k^4 \cosh^2{\omega_k(\tau-\tau_0)}}{k^2\omega_k^4
    + 2k^2\omega_k^2(\omega_k^2-m^2)\sinh^2{\omega_k(\tau-\tau_0)}
    +(\omega_k^2-m^2)^2\sinh^4{\omega_k(\tau-\tau_0)}}
    - \epsilon m
\end{align}
depicted in figure \ref{fig:lagrb}. As shown in the figure, the two peaks emerge as explained previously that the (real) bion configuration can be viewed as a combination of a kink and an anti-kink. As $k$ increases, the overlap between two lumps becomes flat and smear out across Euclidean time.
{\noindent}The associated action 
\begin{align}\label{eq:rbaction}
    \srb = \frac{\omega_k}{g^2\sqrt{k^2-1}}
    \left[ a_+^{-1}\log(\frac{a_++1}{a_+-1}) - a_-^{-1}\log(\frac{a_-+1}{a_--1})\right]
\end{align}
where the vacuum value $-\epsilon m$ is neglected and 
\begin{align}
    a_\pm = \frac{\sqrt{k} \omega_k}{ \sqrt{km^2 \pm\sqrt{k^2-1}(\omega_k^2-m^2)}}
    \,.
\end{align}
When we perform the path integral calculation around the real bion point in the configuration space, the ground state energy then acquires non-perturbative correction due to the non-trivial action. In particular, the leading order behavior of the action takes the form 
\begin{align}
    \srbz = \frac{2m}{g^2}\frac{\log R_0}{\sqrt{k^2-1}} 
    \qq{with}
    R_0 \equiv k + \sqrt{k^2-1}
    \,.
\end{align}
For the latter convenience in Sec. \ref{sec:epi} and Sec. \ref{sec:nsr}, we denote the ``effective'' mass parameter of the deformed $\cp^1$ as 
\begin{align}
    \ms = \frac{m\log R_0}{\sqrt{k^2-1}} \,.
\end{align}

\subsubsection*{Complex bion solution}
In \cite{Behtash:2015loa,Fujimori:2016ljw,Fujimori:2017oab}, some substantial evidences are provided that the complexified solutions play an important role in the study of the resurgence structure of a quantum mechanical system.
For this purpose, we can complexify the configuration space and search for the solution to the equation of motion \eqref{eq:saddle1}. That is, we allow $\theta(\tau)$ in \eqref{eq:ellpara} to be complex. Note that upon the complexification, $\bphi{}$ reveals itself as an independent variable from $\varphi$ and is no longer the complex conjugate of $\varphi$. To avoid ambiguity, we adopt the notation $\tphi$ for the complexified $\bphi{}$.

Based on the real bion solution, the complex bion solution then takes the form
\begin{align}
\label{eq:cpxbion}
\begin{aligned}
    \varphi_{\rm cb}(\tau) &= e^{i\alpha}\sqrt{\frac{k\omega_k^2}{\omega_k^2-m^2}}
    \frac{1}{i\cosh{\omega_k(\tau-\tau_0)}}
    \\[1mm]
    \tphi_{\rm cb}(\tau) &= e^{-i\alpha}\sqrt{\frac{k\omega_k^2}{\omega_k^2-m^2}}
    \frac{1}{i\cosh{\omega_k(\tau-\tau_0)}}
\end{aligned}    
\end{align}
As noted in \cite{Fujimori:2016ljw}, this solution can also be obtained by shifting $\tau_0$ in the real bion solution with an imaginary part $\tau_0 \to \tau_{0} + \frac{1}{\omega_k}\frac{i\pi}{2}$. 
\begin{figure}[t] 
    \centering 
    \begin{subfigure}[b]{.47\linewidth}
      \centering
      \includegraphics[width=.9\textwidth]{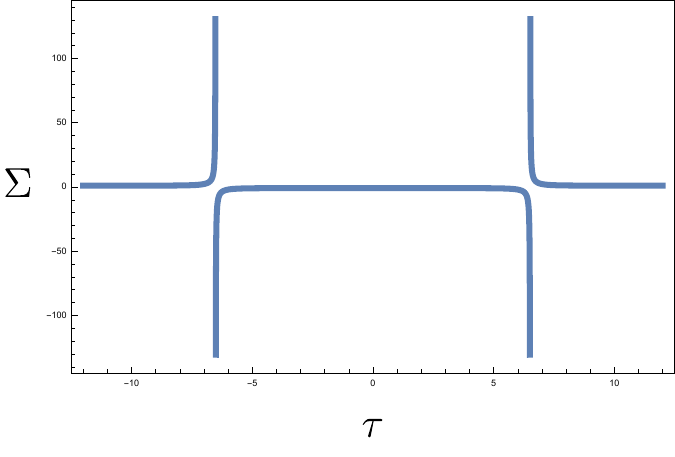}
      \caption{\small $k=2$}
    \end{subfigure}
    \begin{subfigure}[b]{.47\linewidth}
      \centering
      \includegraphics[width=.9\textwidth]{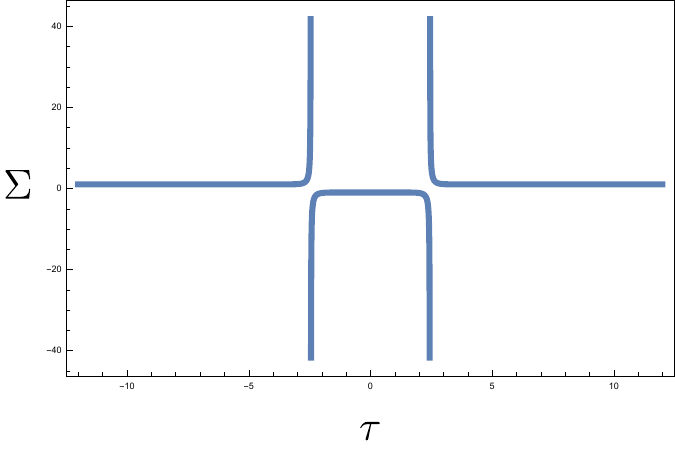}
      \caption{\small $k=10^6$}
    \end{subfigure}
    \caption{\small The profile function of the complex bion solution. The parameters are given as follows. $m=1, \epsilon=1.1, g=1/500$.}
    \label{fig:cbprofile}
\end{figure}
\begin{figure}[H] 
    \centering 
    \begin{subfigure}[b]{.47\linewidth}
      \centering
      \includegraphics[width=.9\textwidth]{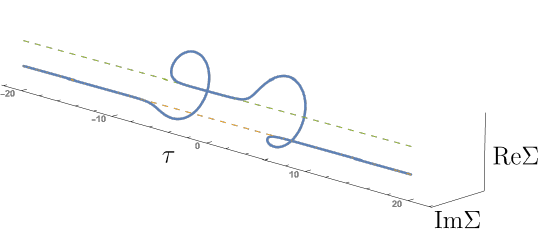}
      \caption{\small $k=2$}
    \end{subfigure}
    \begin{subfigure}[b]{.47\linewidth}
      \centering
      \includegraphics[width=.9\textwidth]{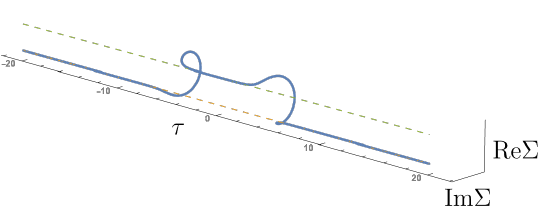}
      \caption{\small $k=10^4$}
    \end{subfigure}
    \caption{\small The regularized profile function of the complex bion solution. The parameters are given as follows. $m=1, \epsilon=1.1, g=e^{0.01i}/500$.}
    \label{fig:cbrprofile}
\end{figure}
The complex bion solution also admits a representation as a kink-antikink pair, say,
\begin{align}
\label{eq:cbkak}
    \begin{aligned}
        \varphi_{\rm cb} &= \left[  
            e^{\omega_k(\tau - \tilde{\tau}_+) - i\tilde{\alpha}_+} + e^{-\omega_k(\tau - \tilde{\tau}_-) - i\tilde{\alpha}_-}
        \right]^{-1}
        \\ 
        \tphi{}_{\rm cb} &= \left[  
            e^{\omega_k(\tau - \tilde{\tau}_+) + i\tilde{\alpha}_+} + e^{-\omega_k(\tau - \tilde{\tau}_-) + i\tilde{\alpha}_-}
        \right]^{-1}
    \end{aligned}
\end{align}
where the complexified positions and phases of kink and anti-kink, respectively, are 
\begin{align}\label{eq:cbr}
    \tilde{\tau}_{\pm} = \tau_0 \pm \frac{1}{2\omega_k} \left( \log\frac{4k\omega_k^2}{\omega_k^2-m^2} -i\pi \right)
    \qand
    \tilde{\alpha}_{\pm} = \alpha
    \,.
\end{align}
The profile function is visualized in figure \ref{fig:cbprofile}. Unlike the real bion configuration, the complex bion diverges and has to be regularized.
This is also indicated in the Lagrangian of the complex bion 
\begin{align}
    \cL_{\rm cb} = -\frac{4m\epsilon k^2\omega_k^4 \sinh^2{\omega_k(\tau-\tau_0)}}{k^2\omega_k^4
    - 2k^2\omega_k^2(\omega_k^2-m^2)\cosh^2{\omega_k(\tau-\tau_0)}
    +(\omega_k^2-m^2)^2\cosh^4{\omega_k(\tau-\tau_0)}}
    - \epsilon m
\end{align}
where the poles are saturated at 
\begin{align}\label{eq:cbpole}
    \tau_{\ast}^{\pm\pm} = \tau_0 \pm \frac{1}{\omega_k}\arc\cosh(\sqrt{\frac{k\omega_k^2(k\pm\sqrt{k^2-1})}{\omega_k^2-m^2}})
\end{align}
Note that the pole structure of the complex bion in the deformed model is quite different from the $\cp^{1}$ case (i.e. $k=1$). In the deformed model, there are four simple poles where two of them are greater than $\tau_0$ and the others are smaller than $\tau_0$ on the real line. As we take the limit $k \to 1$, each pair collapses to a single pole.

To regularize the configuration and the corresponding integral, we can turn on a small imaginary part of the coupling constant $g^2$ to smooth out the singular points. With the complexified $g^2$, we can remove the discontinuities in the complex bion profile, but the price to pay is that the profile becomes complex, see figure \ref{fig:cbrprofile}.

\begin{figure}[t] 
    \centering 
    \begin{subfigure}[b]{.47\linewidth}
      \centering
      \includegraphics[width=\textwidth]{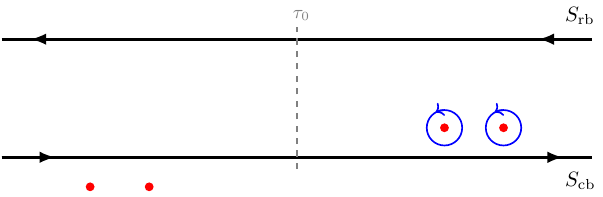}
      \caption{\small $\arg(g^2)<0$}
    \end{subfigure}
    \begin{subfigure}[b]{.47\linewidth}
      \centering
      \includegraphics[width=\textwidth]{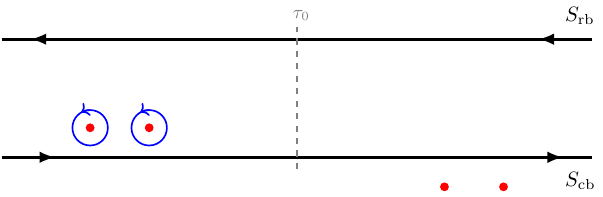}
      \caption{\small $\arg(g^2)>0$}
    \end{subfigure}
    \caption{\small The integration contour of the difference between $S[\varphi_{\rm rb}]$ and $S[\varphi_{\rm cb}]$ over the complex variable $\tau$. The sign of the complexified coupling $g^2$ determines which pair of poles is include in the contour.}
    \label{fig:cbpoleg}
\end{figure}

As previously mentioned, the complex bion deviates from the real bion solution by a constant shift in $\tau_0$ of $i\pi/2\omega_k$, which can be regarded as shifting the complex bion's integration path by $-i\pi/2\omega_k$.
In other words, we return to the same integral of $\srb$ except for changing the integration contour from $(-\infty, \infty)$ to $(-\infty -  \frac{1}{\omega_k}\frac{i\pi}{2}, \infty- \frac{1}{\omega_k}\frac{i\pi}{2})$. See figure \ref{fig:cbpoleg} for the plots of the contours.
It then suffices to consider the difference between $\srb$ and $S[\varphi_{\rm cb}]$ by calculating the residues around the poles \eqref{eq:cbpole},
\begin{align}\label{eq:cbrbdiff}
    S[\varphi_{\rm cb}] - \srb 
    &= 2\pi i \sum_{\tau_{\ast}^{\pm\pm} \in C}\Res[\cL_{\rm rb}]
    \nonumber\\ 
    &=
    \pm 2\pi i \cdot \frac{2\epsilon\sqrt{m}}{\sqrt{m-2g^2\epsilon\sqrt{k^2-1}} + \sqrt{m+2g^2\epsilon\sqrt{k^2-1}}}
\end{align}
where $C$ is the union of the integration contour of the real and complex bions on the complex $\tau$ plane. The plus and minus signs in Eq. \eqref{eq:cbrbdiff} depend on the choice of the sign of $\arg(g^2)$. To be more precise, the contour integration picks up the residues at $\tau_{\ast}^{-\pm}$ and $\tau_{\ast}^{+\pm}$ when $\arg(g^2)>0$ and $\arg(g^2)<0$, respectively. In our discussion, we focus on the situation $g^2$ is small enough such that $2g^{2}\epsilon\sqrt{k^2-1}<m$.
Up to the one-loop level $\order*{g^2}$, the difference turns out to be 
\begin{align}\label{eq:dcrol}
    S[\varphi_{\rm cb}] - \srb  = \pm 2\pi i \cdot \biggl[\epsilon + \order*{g^4}\biggr]
    \,,
\end{align}
coinciding with the case of the $\cp^1$ model.

\subsubsection{Multibion configuration}

In the previous discussion, we have recognized the existence of the leading-order saddle point configurations. However, this is not the end of the story. The ground state energy can still receive the contributions from the $p$-bions as realized in the path integral formalism \cite{Fujimori:2017osz,Fujimori:2017oab}. We provide a summary of essential properties of multibion in the deformed $\cp^1$ model for subsequent use in Sec. \ref{sec:nsr}. A more detailed description can be found in Appendix \ref{sec:mbs}.

\begin{figure}[t] 
    \centering 
    \begin{subfigure}[b]{.47\linewidth}
      \centering
      \includegraphics[width=.9\textwidth]{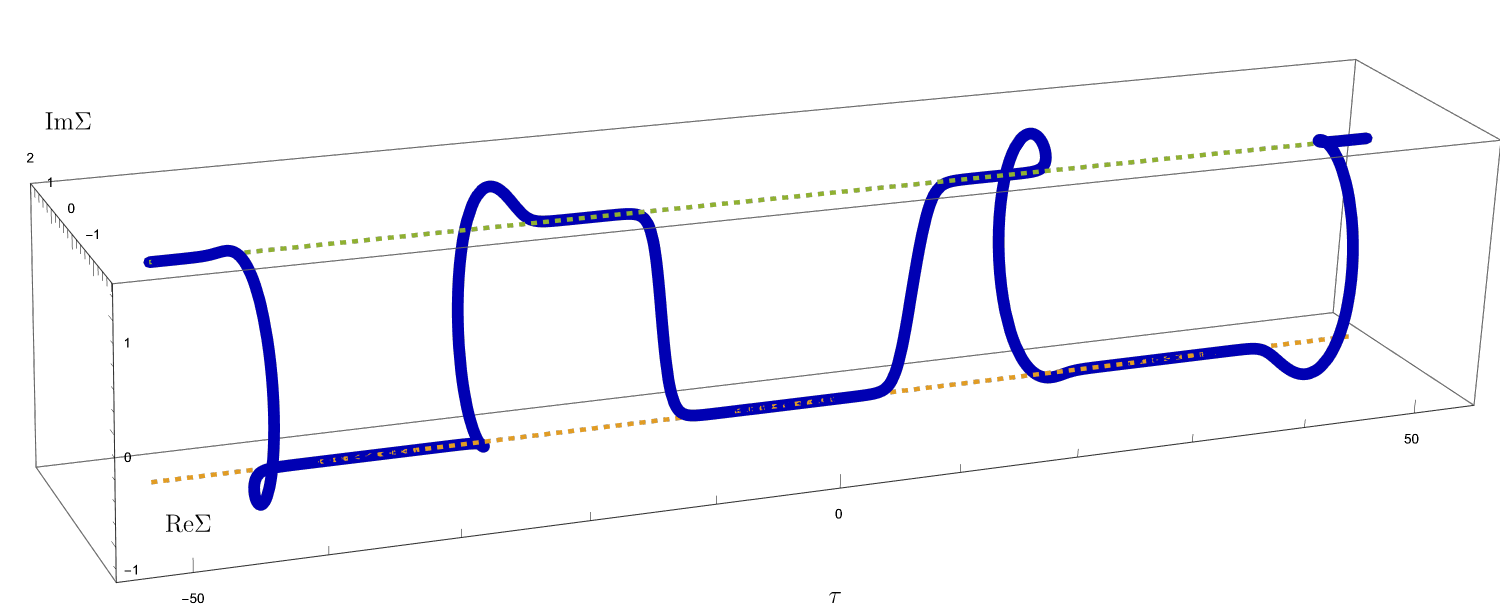}
      \caption{\small $p=3$, $q=1$}
    \end{subfigure}
    \begin{subfigure}[b]{.47\linewidth}
      \centering
      \includegraphics[width=.9\textwidth]{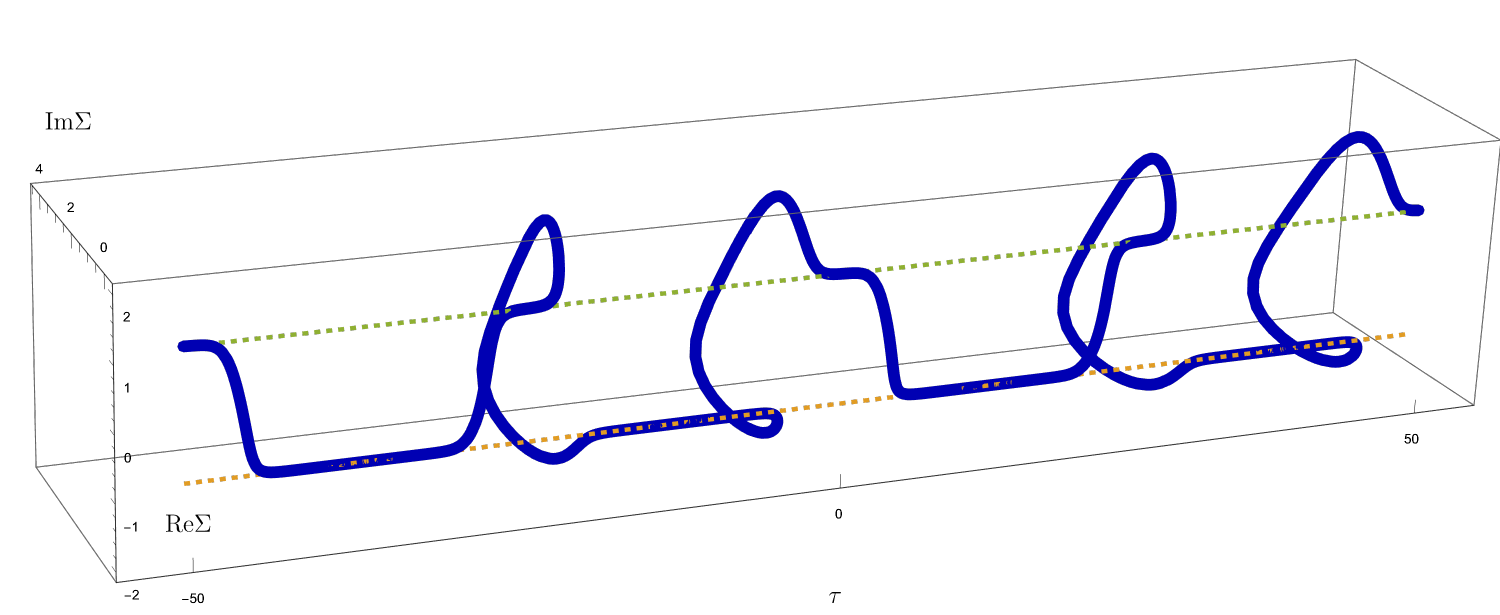}
      \caption{\small $p=4$, $q=1$}
    \end{subfigure}
    \caption{\small The regularized profile function of the multibion solution. The parameters are given as follows. $m=1, k=2, \epsilon=1.1, g=e^{0.01i}/200$, $\beta=100$. The $x,y,z$ axes are the Euclidean time $\tau$, the real part of $\Sigma$, and the imaginary part of $\Sigma$, respectively.} 
    \label{fig:mltbion}
\end{figure}

Along the same line of the consideration in \cite{Fujimori:2017oab}, let us first compactify the remaining time dimension as a circle $S^1_{\beta}$ and introduce the periodic boundary condition of the complex field, $\varphi(\tau) = \varphi(\tau+\beta)$.
For the time being, we assume the energy in \eqref{eq:EandJ} to be undetermined. Take the ansatz 
\begin{align}\label{eq:mbansatz}
    \varphi = e^{i\alpha}f(\tau-\tau_0)
    \,,\quad
    \tphi = e^{-i\alpha}f(\tau-\tau_0)
    \,,
\end{align}
the solution to Eq. \eqref{eq:EandJ} is 
\begin{align}\label{eq:mbphi}
    \varphi(\tau) = \frac{e^{i\alpha}}{\lambda} \cs(\Omega_k(\tau-\tau_0),\xi_k^2)
\end{align}
and $\tphi$ with an opposite phase factor. In Eq. \eqref{eq:mbphi}, $\cs(z,a)$ is a Jacobi elliptic function, $\lambda$ is a free parameter, and $\Omega_k$ as well as $\xi_k$ can then be parametrized as 
\begin{align}\label{eq:mbpara1}
    \begin{aligned}
        \Omega_k^2 &= \frac{(-k+\lambda^2)\,\omega_k^2}{k(1-2k\lambda^2+\lambda^4)} + \frac{m^2(2k-\lambda^2)}{k(1-2k\lambda^2+\lambda^4)}
        \\ 
        \xi_k^2 &= \frac{m^2\lambda^2}{\Omega^2(k-\lambda^2)} + \frac{k-2\lambda^2+k\lambda^4}{k-\lambda^2}
    \end{aligned}
\end{align}
with the energy 
\begin{align}\label{eq:mbpara2}
    E 
    = \frac{(1-\lambda^4)\,\Omega_k^2-m^2}{2g^2(k-\lambda^2)}
    \,.
\end{align}
Alternatively, $\Omega_k, \xi_k$, and $\lambda$ can be associated to the period $\beta$ of the compactified dimension, 
\begin{align}\label{eq:period}
    \beta = 
    \frac{1}{\Omega_k}\left[ 2pK(\xi_k^2) + 4qiK\left( 1-\xi_k^2 \right) \right]
\end{align}
where $K(z)$ is the complete elliptic integral of the first kind, and $p, q$ are integers determining the type of bions.
Multibion solutions under the large $\beta$ limit are depicted in figure \ref{fig:mltbion}. Because the multibion solution diverges at the turning points as the complex bion, we again regularize the profile function by a small phase of the coupling constant $g^2$.
One can immediately find from figure \ref{fig:mltbion} that indeed the integer $p$ is the number of equally spacing bions within a period $\beta$.

As we are interested in the physics of the flat time dimension, i.e. the large $\beta$ limit, it suffices to approximate $\Omega_k, \xi_k$, and the energy as 
\begin{align}\label{eq:paramb}
\begin{aligned}
    \xi_k &\approx 1-8\exp(-\frac{\omega_k\beta-2\pi iq}{p})
    \\[1mm] 
    \Omega_k & \approx \omega_k + 8\omega_k \frac{(2k^2-1)\omega_k^2+m^2}{\omega_k^2-m^2}\exp(-\frac{\omega_k\beta-2\pi iq}{p})
    \\[1mm]
    E &\approx m\epsilon + \frac{16k\omega_k^2}{g^2}\frac{\omega_k^2}{\omega_k^2-m^2}\exp(-\frac{\omega_k\beta-2\pi iq}{p})
    \,.
\end{aligned}
\end{align}
As we take the limit $\beta \to \infty$, it is evident that we return to the one-dimensional space, and the values of $\Omega_k$ and $E$ approach $\omega_k$ and $m\epsilon$, respectively. In addition, by neglecting the vacuum constant in the action, one can find the action of the multibion configuration has the form 
\begin{align}\label{eq:lbaction}
    S_{\rm bion}
    = p \srb
    + l \Bigl(S[\varphi_{\rm cb}] - \srb\Bigr) 
\end{align}
in which $l$ is an integer. The latter term is imaginary due to the difference between the complex and the real bion action as shown in Eq. \eqref{eq:cbrbdiff}.
As $\beta$ is sufficiently large, a $p$-bion is constituted by $p$ individual bions equally aligned on a large circle of circumference $\beta$.

\section{Ground state energy from path integral}
\label{sec:epi}

After the discussion of multibion configurations in the last section, we start to look into how they contribute to the corrections of the ground state energy. That is, we study such the correction in the limit, i.e. finite $\epsilon$ and $g^2 \to 0 $ via the path integral formulation. This result can be matched with the result from the standard perturbation theory which will be shown in the next section.

Following the standard lore of quantum mechanics, the thermal partition function can be obtained by considering a path integral over a compactified time direction with the circumference $\beta$. Namely,
\begin{align*}
    Z(\beta) 
    = \Tr e^{-\beta \hat{H}}
    = \int_{\varphi(\tau) = \varphi(\tau+\beta)} \cD{\varphi}\; \exp^{-S[\varphi]}
    \,
\end{align*}
in which the integral is performed over the periodic configurations $\varphi(\tau) = \varphi(\tau+\beta)$.
The ground state energy can be derived by taking the large $\beta$ limit i.e.
\begin{align}
    \tilde{E} = - \lim_{\beta \to \infty} \frac{1}{\beta} \log{Z} \,.
\end{align}
In particular, under the weak coupling limit $g^2 \to 0$, it suffices to consider the following saddle point expansion, 
\begin{align}\label{eq:piEt}
    \tilde{E} = \left( -\lim_{\beta \to \infty} \frac{1}{\beta} \log{Z_0} \right)
    - \lim_{\beta \to \infty} \frac{1}{\beta} \log( 1 + \sum_{p=1}^{\infty} \frac{Z_p}{Z_0})
\end{align}
where $Z_0$ is the contribution from the trivial vacuum while $Z_p$ is the $p$-th leading saddle point. In our case, $Z_p$ can be related to the $p$-bion contribution which can be obtained by utilizing the Lefschetz thimble method.

\subsection{One-loop corrections}

As pointed out in \cite{Fujimori:2016ljw,Fujimori:2017oab,Fujimori:2017osz}, one has to take both zero modes and quasi zero modes into account when computing the one-loop corrections around the saddle points. If only the zero modes are isolated in the path integral, the resulting modification to the ground state energy is valid exclusively in the supersymmetric limit.\footnote{That is, the one-bion correction to the ground state energy under the Gaussian approximation takes the form 
\begin{align}\label{eq:onebiongapprox}
    \tilde{E}_{1}^{\times} = -\frac{16ik\omega_k^4}{g^2(\omega_k^2-m^2)} \left( 1 - e^{\Delta S_{\rm cr}} \right)e^{-\srb}
\end{align}
in which $\Delta S_{\rm cr}$ is the difference between the real and complex bion actions, c.f. Eq. \eqref{eq:cbrbdiff}. Then up to the one-loop level, we can see from Eqn. \eqref{eq:dcrol} that $E_1^{\times}$ matches with our expectation that the nonperturnative contribution vanishes in the supersymmetric regime.
}
To incorporate the effect of quasi zero modes, let us first look into the solution to the valley equation \cite{Balitsky:1986qn,Yung:1987zp},
\begin{align}\label{eq:valley}
    \left.\fdv{S}{\varphi}\right|_{\varphi=\varphi_B} =  
    GK^{\nu}\pdv{\tphi{}_B}{\eta^{\nu}}
    \qand
    (\varphi_B \xleftrightarrow{~~~} \tphi{}_B)
    \,,
\end{align}
where $\eta^{\mu} = (\tB{r},\aB{r})$ and $K^{\nu}$ is some coefficient associated to the metric of the moduli space. Eq. \eqref{eq:valley} can be solved, up to the leading order $\order*{g^2}$, by the kink-antikink ansatz\footnote{
    In the weak coupling limit, the valley equation \eqref{eq:valley} is verified up to the order $\order{g^2}$ with the coefficient 
    \begin{align}
        K^{\tB{r}} = \frac{4k}{m} \pdv{S_{\rm eff}}{\tB{r}}
        \qand
        K^{\aB{r}} = 4km \cdot \pdv{S_{\rm eff}}{\aB{r}} \,.
    \end{align}
    Also, it is straightforward to check that $K^{\nu}$ vanishes at the saddle points, i.e. the real and complex bions.
}of which the fields take the form\footnote{Note that as $\tau_{B\pm}$ and $\alpha_{B\pm}$ take values of $\tau_{\pm}, \alpha_{\pm}, \tilde{\tau}_{\pm}, \tilde{\alpha}_{\pm}$, $\varphi_B$ and $\tphi{}_B$ are then the real and complex bion solutions, respectively. In other words, we will consider the integration over the quasi-moduli space $(\tB{r},\aB{r})$ along the contour passing through the real (and complex) bion saddle, i.e., $\cJ_{\rm rb/cb}$ or $\cK_{\rm rb/cb}$.}
\begin{subequations}
    \label{eq:kkansatz}
    \begin{align}
        \varphi_{B} &= \left[  
                e^{\omega_k(\tau - \tB{+}) - i\aB{+}} + e^{-\omega_k(\tau - \tB{-}) - i\aB{-}}
            \right]^{-1}
            \\ 
        \tphi{}_{B} &= \left[  
            e^{\omega_k(\tau - \tB{+}) + i\aB{+}} + e^{-\omega_k(\tau - \tB{-}) + i\aB{-}}
            \right]^{-1} \,,
    \end{align}
\end{subequations}
where $\tB{r} = \tB{+}-\tB{+}$ and $\aB{r} = \aB{+}-\aB{+}$. The asymptotic behavior of the bion Eq. \eqref{eq:kkansatz} looks like 
\begin{align*}
    \varphi_B = \left\{\begin{aligned}
        & e^{m(\tau-\tB{-}) + i \aB{-}}
        - e^{3m(\tau-\tB{-})}e^{-m\tB{r} + i(2\aB{-}-\aB{+})}
        + \order*{g^2}
        \,,&& \tau \approx \tB{-}\\[2mm]
        & \infty
        \,,&& \tau \approx \frac{\tB{-}+\tB{+}}{2} \equiv \tau_{0} \\[2mm]
        & e^{-m(\tau-\tB{+}) + i \aB{+}}
        - e^{-3m(\tau-\tB{+})}e^{-m\tB{r} + i(2\aB{+}-\aB{-})}
        + \order*{g^2}
        \,,&& \tau \approx \tB{+}
    \end{aligned}\right.
\end{align*}

Under one-bion background configuration, the effective action takes the form 
\begin{align} \label{eq:Seffkk}
    S_{\rm eff}(\tB{r},\aB{r}) = 
    \frac{2m}{g^2}\left( \frac{\log{R_0}}{\sqrt{k^2-1}} - 2e^{-m\tB{r}}\cos{\aB{r}} \right)
    + 2m\epsilon\tB{r}
    + \order{g^2}
\end{align}
where the first term in the parentheses is due to the asymptotic behavior of a single kink and anti-kink, the second term is the interaction between a kink-anti-kink pair under large separation, and the last confining term is due to the correlation between fermion zero modes. The effective action \eqref{eq:Seffkk} differs from the undeformed model \cite{Fujimori:2016ljw,Fujimori:2017oab} only by the constant term.

Next, consider the expansion around the kink-anti-kink background $\varphi_B$ with the quantum fluctuation $\delta\varphi$ orthogonal to the quasi-zero modes, say, 
\begin{align}
    \mqty(\varphi \\[2mm] \tphi{}) = 
    \mqty(\varphi_B \\[2mm] \tphi{}_B)
    +
    \frac{gG^{-1/2}_B}{\sqrt{2}} \mqty( \xi_1 + i \xi_2 \\[2mm] \xi_1 - i \xi_2 )
    \,,\qquad
    \int G \pdv{\tphi{}_B}{\eta^{\nu}}\delta\varphi 
    = \int G \pdv{\varphi_B}{\eta^{\nu}}\delta\tphi{}
    = 0 
\end{align}
where $G_B$ denotes the metric under $\varphi_B$ background and $\xi_i$ is the quantum fluctuation.
With such a choice, there is no linear terms due to the valley equation and the action up to the quadratic order reads 
\begin{align}
    S = S_{\rm eff} 
    + \frac{1}{2}\int \xi^{T}\Delta_{B} \xi + \order{g^2}
\end{align}
where $S_{\rm eff}$ is the effective action evaluated at the saddle point. The differential operator of the second order term is 
\begin{align}\label{eq:difflucB}
    \Delta_{B} = -\d_\tau^2 + \tilde{V}_{B}
\end{align}
where $\tilde{V}_B$ is a two-by-two matrix by substituting the kink-anti-kink background into the second order expansion. 

After isolating the zero and quasi-zero modes and integrating over the rest of excitations, we then arrive at
\begin{align}\label{eq:piE1}
-\lim_{\beta \to \infty} \frac{1}{\beta}\frac{Z_1}{Z_0} &= 
-\lim_{\beta \to \infty} \frac{1}{\beta}\int \dd{\tau_0}\dd{\alpha_0}\dd{\tB{r}}\dd{\aB{r}}
\left[ \det(\frac{\cG}{2\pi})\det(\frac{\cG'}{2\pi})\frac{\det\Delta_0}{\det''\Delta_B} \right]^{1/2}
e^{-S_{\rm eff}}
\nonumber\\[2mm]
&= 
-\frac{8m^4}{\pi g^4} 
\int e^{-S_{\rm eff}} \dd{\tB{r}}\dd{\aB{r}}
\,.
\end{align}
Here $\cG$ and $\cG'$ are the metric on the moduli and quasi-moduli spaces, respectively, and $\Delta_0$ is the differential operator around the minimum ($\varphi, \tphi{}=0$.) In the second line, we use the fact that the element in the square bracket can be approximated as a double copy of a single kink for a well-separated kink-anti-kink pair \cite{Fujimori:2016ljw,Fujimori:2017osz}.
Also, the ratio from the collective coordinate is fixed in the way
\begin{align}
    \lim_{\beta \to \infty} \frac{1}{\beta} \int\dd{\tau_0} = 1
    \,.
\end{align}
Note that the prefactor in \eqref{eq:piE1} is identical to the one from the $\cp^1$ model under the kink approximation. This arises because the kink solution in the Lie algebraic deformed $\cp^1$ model is the same as that of the undeformed $\cp^1$ model, as can be observed directly from the explicit formulation \cite{SS2023}. The rest of the task is to perform the integration over the quasi-moduli space via the Lefschetz thmible method (see e.g. \cite{Marino:2012zq,Aniceto:2014hoa,Dorigoni:2014hea,sauzin2014introduction1summabilityresurgence} for recent reviews.) Schematically, the integral takes the form 
\begin{align}
    [\cI\overline{\cI}]
    \equiv
    \int_{-\infty}^{\infty} \dd{\tB{r}} \int_{-\pi}^{\pi} \dd{\aB{r}} \exp(
        \frac{4m}{g^2} e^{-m\tB{r}}\cos{\aB{r}}
        -2m\epsilon\tB{r}
    )
\end{align}
in which the integration contour of $\tB{r}$ and $\aB{r}$ should be realized as the linear combination of the thimbles $\cJ_{\sigma},\cK_{\sigma}$.
To this end, we have to first complexify the quais-moduli variables and deform the integration contour to a middle dimensional path via the flow equation such that the overall path is the same the original one. In other words, the coupling constant is made complex $g^{2}e^{i\theta}$ with a small phase $\theta$, leading to complex $\tB{r}$ and $\aB{r}$. The flow equation is 
\begin{align}\label{eq:floweq}
    \dv{\tB{r}}{t} = \frac{\sqrt{k^2-1}}{2m\log{R_0}}\overline{\pdv{S_{\rm eff}}{\tB{r}}}
    \,,\quad
    \dv{\aB{r}}{t} = \frac{m\sqrt{k^2-1}}{2\log{R_0}}\overline{\pdv{S_{\rm eff}}{\aB{r}}}
\end{align}
Note that this can be recast in the same form of the flow equation of the $\cp^1$ quantum mechanics. See Appendix \ref{sec:apppi} for more detail. 
As a consequence, the first order correction to the ground state energy from the one-bion saddle point is 
\begin{align}\label{eq:pioneb}
\tilde{E}_{1} &=
-\lim_{\beta \to \infty} \frac{1}{\beta}\frac{Z_1}{Z_0} = -2m\left( \frac{g^2}{2m} \right)^{2(\epsilon-1)}\frac{\sin{\epsilon\pi}}{\pi}\Gamma^{2}(\epsilon)\exp(-\operatorname*{sgn}(\theta)\pi i \epsilon)
e^{-\srbz}
\end{align}

In fact, the above result can be further generalized to the case of multibion contributions by following the same strategy in \cite{Fujimori:2017oab,Fujimori:2017osz}. One can immediately obtain the $p$-bion contribution 
by replacing the value of the effective at each saddle point since the part depending on the quasi zero mode in the effective action is equivalent to the $\cp^1$ case as shown in \eqref{eq:Seffkk}.
In broad terms, the expansion of the second-order correction to the ground state energy can be approximated by summing the leading-order term of $Z_p/Z_0$ in $\beta$, consistent with the dilute gas approximation. Namely,
\begin{align}\label{eq:e2sketch}
    \tilde{E}_2 = -\frac{2m\Gamma(\epsilon)}{\Gamma(1-\epsilon)}e^{-\srbz\mp\pi i \epsilon}\left( \frac{2m}{g^2} \right)^{2(1-\epsilon)}
    + \cdots 
\end{align}
The exact formula of the multibion contribution $Z_p/Z_0$ \eqref{eq:pbionZ} is quoted in Appendix \ref{sec:apppi}. 
Also, the further comparison between the current result and the one from the standard perturbation can be seen in the next section.

To wrap up this section, the one-bion correction is presented in \eqref{eq:pioneb} while the ingredient to get the multibion one is shown in \eqref{eq:pbionZ} from which one can derive the associated modification together with \eqref{eq:piEt}.
Moreover, we point out that the ambiguities in \eqref{eq:pioneb} and \eqref{eq:pbionZ} arise from the sign of $\theta$ at the one-loop level when $\epsilon$ deviates from unity. However, at the leading order $(\epsilon-1)$, no ambiguity is present; it only arises at the next-to-leading order $(\epsilon-1)^2$ as we will see below.

\section{Nearly supersymmetric regime}
\label{sec:nsr}

In this section, we examine the correction to the ground state energy for finite $g^2$ and $\epsilon \to 1$. Specifically, we use Rayleigh-Schrödinger perturbation theory to calculate the perturbative correction introduced by the deformed potential near the supersymmetric point, yielding an exact result within the $g^2$ series. Additionally, we compare results obtained from two limiting cases in the weak coupling and nearly supersymmetric regime.

\subsection{$\delta\epsilon$ perturbation}

As we have seen in the previous analysis of Sec. \ref{sec:model} that in the supersymmetric case $\epsilon=1$, the (supersymmetric) ground state energy is exactly zero. Now, we would like to further investigate the correction to the ground state energy due to the deformed boson-fermion interaction with $\epsilon\neq 1$. To be more precise, we first expand in 
\begin{align}
    \delta\epsilon = \epsilon-1    
\end{align}
such that the eigenenergy takes the form 
\begin{align}
    E 
    = \sum_{n=1}^{\infty} (\delta\epsilon)^n E^{(n)}
    = E^{(1)} \delta\epsilon + E^{(2)} (\delta\epsilon)^2 + \cdots
\end{align}
with the eigenfunction reads
\begin{align}
    \Psi = \Psi^{(0)} + \Psi^{(1)} \delta\epsilon + \cdots
    = e^{-\mu/g^2} + \Psi^{(1)} \delta\epsilon + \cdots
\end{align}

To begin with, by solving the first order Schr\"odinger equation \eqref{eq:hgrd}, one can find the ground state energy is modified in the way
\begin{align}\label{eq:eeone}
    E^{(1)} &= -\frac{1}{\braket{0}{0}}
    \ev**{\frac{1}{G}\pdv{\mu}{\varphi}{\bphi{}}}{0}
    = \frac{m^2}{g^4(k^2-1)-m^2}\frac{g^2k+m - (g^2k-m)R_0^{\frac{2m}{g^2\sqrt{k^2-1}}}}{-1 + R_0^{\frac{2m}{g^2\sqrt{k^2-1}}}}
\end{align}
(The detailed calculation is given in Appendix \ref{sec:pertb1}.)
From Eq. \eqref{eq:eeone}, we have found the signal of the trans-series behavior. Indeed, notice that $R_0^{-2m/g^2\sqrt{k^2-1}}$ can be rewritten as 
\begin{align}\label{eq:epfct}
    R_0^{-\frac{2m}{g^2\sqrt{k^2-1}}} 
    = \exp(-\frac{2m}{g^{2}}\frac{\log{R_0}}{\sqrt{k^2-1}})
\end{align}
which will serve as the indicator\footnote{The non-perturbative weight has the form $\sim e^{-I}$ and can be extracted from $e^{-\srb}$ to the leading order, which only the bosonic part contributes at the one-loop level. In particular, the bosonic sector of the model discussed in this paper can be associated to the $\eta$-deformed model \cite{Delduc:2013fga,Bykov:2020llx} as previously mentioned.
If we identify two parametrizations between $k$ and $\eta$ then two actions of the pure bosonic theory are identical up to an overall constant. The exponential factor is treated as the non-perturbative factor in the similar transseries expansion in \cite{Demulder:2016mja}.
} of the non-perturbative contributions as it is an exponentially small factor. Recall that with the small coupling $g_{\rm 2d}^2$ in Eq. \eqref{eq:comppara}, we focus on the region $g^2 \ll m$ with the additional factor $\log R_0/\sqrt{k^2-1}$ being of order $\order*{1}$ because of $\log R_0/\sqrt{k^2-1} \in [0,1]$ for $k \in [1,\infty)$. Thus we get an exponentially small factor \eqref{eq:epfct} and it returns back to the ordinary $\cp^1$ with the factor $e^{-2m/g^2}$ in the limit $k \to 1$.
In this sense, it suffices to express Eq. \eqref{eq:eeone} as 
\begin{align}\label{eq:eeonepnp}
    E^{(1)} 
    &= \frac{m^2}{m^2-g^4(k^2-1)} 
        \left[ (g^2k-m) 
        + \sum_{p=1}^{\infty} (- 2m) \cdot e^{-p\srbz}
    \right]
\end{align}
in which the first term in the square bracket indicates the perturbative contribution (from the trivial vacuum) while the latter term in the sum is attributed to the non-perturbative one (i.e. the $p$-bion sector). One can observe that the expression of the first order correction \eqref{eq:eeonepnp} is exact in $g^2$ and has \emph{no} ambiguity.

To see this is the case, let us consider the weak coupling limit of \eqref{eq:eeonepnp}. To the one-loop order, we discard higher order terms $\order*{g^2}$, resulting in 
\begin{align}
    \frac{m^2}{m^2-g^4(k^2-1)} \cdot (-2m) e^{-p\srbz}
    = -2m e^{-p\srbz} + \order*{g^4}
    \,.
\end{align}
This matches with the path integral calculation in the $\epsilon \to 1$ limit. Namely, for the one-bion contribution, one can directly see from Eqn. \eqref{eq:pioneb}
\begin{align}
\lim_{\epsilon \to 1}\tilde{E}_1 = -2m
\end{align}
while in the multibion case, one also get the coefficient \eqref{eq:pbion1o} equals to $-2m$ for $p$-bion.
Beyond the one-loop approximation, we conjecture that the two-loop correction has to be similar to that of the ordinary $\cp^1$ model, and the difference proportional to the Lie-algebraic deformation $k-1$ in the perturbation series should start emerging from the three-loop order due to the novel factor $g^4(k^2-1)$.

\subsection{The second order corrections}
In fact, one will find that the ambiguity arises at the second-order correction of the ground state energy as a general feature of quantum mechanical models derived from a $\cN=(0,2)$ two-dimensional chiral model \cite{Fujimori:2017osz,Fujimori:2017oab}. To elaborate this, one may consider the general expression of the next-to-leading order correction to the energy in the perturbation theory, namely,
\begin{align}\label{eq:rs2nde}
    E^{(2)} = -\frac{1}{\braket{0}{0}}
    \ev**{H(\epsilon=1)}{\Psi^{(1)}}
    \,.
\end{align}
Note that the first order correction to the wave function $\Psi^{(1)}$ can be solved by the differential equation 
\begin{align}\label{eq:degfnone}
    H(\epsilon=1)\Psi^{(1)}(\varphi,\bphi{}) 
    = \left(  
        E^{(1)} + \frac{m(1-\abs{\varphi}^4)}{1+2k\abs{\varphi}^2+\abs{\varphi}^4}
    \right)e^{-\mu/g^2}
\end{align}
The exact solution of the first order correction to the wave-function to \eqref{eq:degfnone} can be obtained by integrating eqn. \eqref{eq:D3}. Together with \eqref{eq:rs2nde}, one can find the next-to-leading order modification to the ground state energy shown in the form 
\begin{multline}\label{eq:e2}
    E^{(2)} = E^{(2)}_{{\rm namb},1} 
    - \frac{4m^4e^{-\srbz}\sqrt{k^2-1}}{
        \left[ g^4(k^2-1)-m^2 \right]^2\left( 1 - e^{-\srbz} \right)^3
    }
    \times
    \\[2mm] 
    \times 
    4\int_{0}^{\ms} \,
    \left( \sinh^2{\frac{\mu}{g^2}} \right)
    e^{-\frac{2\mu\sqrt{k^2-1}}{m}}
    \left(  
       \frac{e^{-\srbz} + e^{\frac{6\mu\sqrt{k^2-1}}{m}}}{
        e^{\frac{2\mu\sqrt{k^2-1}}{m}} - 1
    }
    \right)
    \dd{\mu} 
\end{multline}
In \eqref{eq:e2}, $E_{{\rm namb},1}^{(2)}$ is the portion with no potential ambiguity while the ambiguity arises from the rest terms. It is important to note that not all terms in the latter part exhibit ambiguities; therefore, further analysis is required to identify and extract them.

Since we are mostly interested in the suspected ambiguity structure shown up in $E^{(2)}$, let us ignore $E^{(2)}_{{\rm namb},1}$ for the time being and focus on how the ambiguity structure emerges. Here we quote the results as follows and detailed calculations are given in Appendix \ref{sec:pertb1}.
By expanding the second line in \eqref{eq:e2} in $\mu$, one would observe the perturbative series in $g^2$ has the structure
\begin{multline}\label{eq:ambstr}
    \frac{2m\left( 1+e^{-\srbz} \right)}{\sqrt{k^2-1}}
        \int_{0}^{\ms} \frac{\sinh^2(\mu/g^2)}{\mu} \dd{\mu} 
    \\[2mm]
    + \sum_{n=1}^{\infty} 
    \frac{2^{1+n}B_{n}(2)}{\Gamma(n+1)}\left( \frac{m}{\sqrt{k^2-1}} \right)^{-n+1}
    \left[ 1 + (-1)^{n}e^{-\srbz} \right]
    \int_{0}^{\ms} \mu^{n-1}\sinh^2\left( \frac{\mu}{g^2} \right)\dd{\mu} 
\end{multline}
where $B_{n}(2)$ is the Bernoulli polynomial evaluated at $2$.
In the second line of \eqref{eq:ambstr}, the $g^2$ series is Borel summable and indicates no ambiguity\footnote{See Appendix \ref{sec:pertb1} for further details.} will emerge while the first term is non-Borel summable which has a pole on the positive real axis. To be more precise, the ambiguous part of the second order correction to the ground state energy turns out to be 
\begin{align}\label{eq:ambE}
    E_{\text{amb}} =
        -\frac{m^5}{\left[ g^4(k^2-1)-m^2 \right]^2}\frac{\cosh{\frac{\ms}{g^2}}}{\sinh^3{\frac{\ms}{g^2}}} \Biggl[
    \pm \frac{i\pi}{2}
    + \frac{1}{2}\int_{0}^{\infty} e^{-t}
        \frac{e^{2\ms/g^2}}{t-\frac{2\ms}{g^2 \pm i0}}
     \dd{t}
    \Biggr]
\end{align}
In the weak coupling regime, we can recast \eqref{eq:ambE} in a trans-series form
\begin{align}
    E_{\rm amb} = \sum_{p=0}^{\infty} E_{p}^{(2)} e^{-2p\ms/g^2}
\end{align}
where the (ambiguous part of) coefficient of the $p$-bion sector is 
\begin{subequations}
\begin{align}
E_0^{(2)} &= \frac{m^5}{\left[ g^4(k^2-1)-m^2 \right]^2} \cdot \int_{0}^{\infty} \frac{e^{-t}}{t-\frac{2\ms}{g^2 \pm i0}}
\dd{t}
\\[2mm]
E_{p>0}^{(2)} &= \frac{m^5}{\left[ g^4(k^2-1)-m^2 \right]^2} \left[  
    \pm 2\pi i p^2 
    +  \int_{0}^{\infty} \frac{(p+1)^2e^{-t}}{t-\frac{2\ms}{g^2 \pm i0}}
    \dd{t}
\right]
\end{align}
\end{subequations}
Here we can already see that the ambiguity cancel out between $E_{p}$ and $E_{p+1}$. Note that the ambiguity structure is of the similar type discovered in \cite{Fujimori:2017oab}.

To the leading order in $g^2$, the $p$-bion background leads to the modification
\begin{align}
    E^{(2)}_{p} = \pm 2\pi i mp^2 + \order{g^2}
\end{align}
The similar phenomenon is observed from the path integral perspective. Indeed, Eqn. \eqref{eq:pbion2o} implies that the same ambiguity appears
while the $g^2$ perturbation series reveals more complicated structure around each saddle. In a comparable manner, since the prefactor of each coefficient corresponds to a geometric series in $g^4(k^2-1)$, additional deviations between the deformed and undeformed models may emerge beyond the two-loop level from the path integral perspective. This is compatible with the aforementioned observation in the first order $\delta\epsilon$ calculation.

\section{Conclusions and outlooks}
\label{sec:clsn}

In this paper, we extend the work of \cite{Fujimori:2016ljw, Fujimori:2017oab} from conventional $\cp^1$ quantum mechanics to a one-parameter deformed model that preserves a specific $U(1)$ symmetry. This deformation is not arbitrary but instead arises naturally from the structure of a two-dimensional Lie-algebraic sigma model, as discussed in recent studies \cite{Sheu:2023hoz, SS2023, Gamayun:2023sif, Gamayun:2023atu}.

In particular, we investigate the nonperturbative effects arising from saddle-point configurations in Lie-algebraically deformed $\cp^1$ quantum mechanics. Beginning with the two-dimensional parent theory containing $N_f$ fermions, the effective theory is derived by compactifying this sausage-like sigma model on $\RR \times S^1$ via the Kaluza-Klein formalism. 
To underscore a significant feature of this model, we present and analyze the saddle-point configurations that play crucial roles in the nonperturbative dynamics. These configurations include the real bion, complex bion, and multibion solutions. 
Fig. \ref{fig:rbprofile} and \ref{fig:cbprofile} illustrate the real and complex bion solutions, respectively, emphasizing on the spatial profiles of these configurations. Fig. \ref{fig:mltbion} further provides a visualization of the multibion configuration, capturing the intricate collective dynamics of multiple bion constituents.
As depicted in Fig. \ref{fig:cbpoleg}, the singularities associated with the complex bion are lifted by the deformation parameter.

Furthermore, the study adopts two complementary perspectives, the path integral formalism and the Rayleigh-Schr\"odinger perturbation theory. The path integral formalism characterizes behavior around different saddle points, which provides the insight of into the model's non-perturbative structure. 
The latter approach is applied to examine the dynamics near the supersymmetric point, capturing the aspects of higher order terms in $g^2$ of the theory. Both approaches align within an overlap regime, specifically at the one-loop level and to second order in $\delta \epsilon$.
At this level, an inherent ambiguity arises due to the shared structural features of the $\cp^1$ model, manifesting particularly at $\delta \epsilon^2$, as discussed in \cite{Fujimori:2016ljw, Fujimori:2017oab}. Through a detailed analysis of the perturbation series, we propose that a Lie-algebraic deformation term, proportional to $(k-1)$, emerges at the three-loop level. This term is driven by a novel factor, $g^4(k^2 - 1)$, suggesting a deeper interplay between the deformation parameter and loop contributions.

This paper concentrates on the quantum mechanical aspects of the deformed $\cp^1$ model. An intriguing direction for future research involves extending these findings to the original two-dimensional formulation. Specifically, applying the Lefschetz thimble method to the deformed $\cp^1$ model in two distinct settings, on $\mathbb{R} \times S^1$, as explored in \cite{Fujimori:2018kqp}, and on $T^2$, offers a suitable approach for probing the model's non-perturbative dynamics from different saddle point configurations. 
Furthermore, given the close relationship between the deformed $\mathbb{CP}^1$ model and the $\eta$-deformed integrable model, it becomes interesting to explore the integrability conditions through a Lie-algebraic framework. 
Such an analysis could reveal features that enhance our understanding of the model's resurgent structure. Utilizing the integrable nature, one can employ the thermodynamic Bethe ansatz to systematically examine these resurgent aspects, providing a pathway to investigate the non-perturbative behavior of the model \cite{Marino:2021six, DiPietro:2021yxb, Marino:2021dzn, Marino:2022ykm}.

\section*{Acknowledgments}

The author is grateful to M. Shifman for helpful discussions and comments.
This work is supported in part by the US National Science Foundation under grant PHY-2210283.

\section*{Appendices}
\appendix

\section{Conventions and special functions}
\label{sec:nc}


\subsection*{Elliptic integrals and elliptic functions}
The (multi)bion solutions and hence its action are certain combinations of elliptic integrals and elliptic functions. Here we provide their definitions and essential properties to derive the actions of different configurations. Further properties can be found in \cite{Byrd_1971,NIST:DLMF}. The definitions of elliptic integrals are given below. 
\begin{align}
    \label{eq:incompef}
    \begin{aligned}
        F(\phi,\xi) &\equiv \int_{0}^{\phi} \frac{\dd{t}}{\sqrt{1-\xi\sin^2{t}}}
        \,,\qquad
        E(\phi,\xi) \equiv \int_{0}^{\phi} \sqrt{1-\xi\sin^2{t}} \dd{t}
        \\[2mm] 
        \Pi(a,\phi,\xi) &\equiv \int_{0}^{\phi} \frac{1}{\sqrt{1-\xi\sin^2{t}}}\frac{1}{1-a\sin^2{t}} \dd{t}
    \end{aligned} 
\end{align}
For some specific value of $\phi$, Eq. \eqref{eq:incompef} reduce to the complete elliptic functions 
\begin{align}
    F\left( \frac{\pi}{2},\xi \right) = K(\xi) \,,\quad
    E\left( \frac{\pi}{2},\xi \right) = E(\xi) \,,\quad
    \Pi\left( a,\frac{\pi}{2},\xi \right) = \Pi(a,\xi) \,.
\end{align}
The correct analytic continuation of elliptic integrals is crucial in the present calculations. To make it, some useful transformation of the argument and the modulus \cite{Byrd_1971} are listed below 
\begin{align}
    F(\phi,\xi) &= iF(\psi,1-\xi)
    \\[2mm]
    E(\phi,\xi) &= i\left[  
        F(\psi,1-\xi) - E(\psi,1-\xi) + \frac{(1-\xi)\sin{2\psi}}{2\sqrt{1-(1-\xi)\sin^2{\psi}}}
    \right]
    \\[2mm] 
    \Pi(a,\phi,\xi) &= i\left[ F(\psi,1-\xi) + \frac{a}{1-a}\Pi\left( \frac{1-\xi}{1-a},\psi,1-\xi \right) \right]
\end{align}
where 
\begin{align*}
    \sin\psi = \frac{\sqrt{\sin{\phi}^2-1}}{(1-\xi)\sin{\phi}}
    \,.
\end{align*}
We will also use the identity
\begin{align}\label{eq:jpiid}
    \lim_{\xi_k^2 \to 1} a \Pi(1+a,\xi_k^2) + b \Pi(1-b,\xi_k^2)
    = \sqrt{1+a} \log\frac{1+\sqrt{1+a}}{1-\sqrt{1-a}}
    - \sqrt{1-b} \log\frac{1+\sqrt{1-b}}{1-\sqrt{1-b}} \,.
\end{align}

Next, the Jacobi elliptic functions are defined as 
\begin{align}\label{eq:elpf}
\begin{aligned}
    \sn(z,\kappa) \equiv \sin(F^{-1}(z,\kappa))
    \,,\quad
    \cn(z,\kappa) \equiv \cos(F^{-1}(z,\kappa))
\end{aligned}
\end{align} 
and 
\begin{subequations}
\begin{align}
    &\dn(z,\kappa) \equiv \sqrt{1 - \kappa\sn^2(z,\kappa)}
    \,,\quad
    \sd(z,\kappa) \equiv \frac{\sn(z,\kappa)}{\dn(z,\kappa)}
    \,,\quad
    \\[2mm]
    &\cd(z,\kappa) \equiv \frac{\cn(z,\kappa)}{\dn(z,\kappa)}
    \,,\quad
    \cs(z,\kappa) \equiv \frac{\cn(z,\kappa)}{\sn(z,\kappa)}
\end{align}
\end{subequations}
which will be useful in describing the multibion solution and the profile functions.

\subsection*{Bernoulli polynomial}

Here, we present the definition and key identities of Bernoulli polynomials, which will be used in discussing the Borel summability of the second-order correction in Appendix \ref{sec:pertb1}.

Consider the following generating function 
\begin{align}
G_B(t,x) 
= \frac{te^{xt}}{e^t-1}
= \sum_{n=0}^{\infty} B_n(x) \frac{t^n}{n!}
\end{align}
where $B_n(x)$ is the Bernoulli polynomial with the expansion 
\begin{align}
    B_n(x) = \sum_{k=0}^{n} B_{n-k}x^k
\end{align}
where $B_{n-k}$ is the Bernoulli number. We will also need the asymptotic behavior of $B_n(x)$ i.e. 
\begin{align}\label{eq:berasym}
B_{n}(x) \sim \frac{2(-1)^{\lfloor{\frac{n}{2}}\rfloor-1}}{(2\pi)^{n}} \Gamma(n+1)
\times
\left\{\begin{aligned}
    &\cos(2\pi x) \,,
    && n \in 2\ZZ 
    \\ 
    &\sin(2\pi x) \,,
    && n \in 2\ZZ +1
\end{aligned}\right.
\end{align}
as $n \to \infty$.

\section{Multibion solutions}
\label{sec:mbs}

In this section, we give a summary of some essential properties of multibion solutions. To begin with, with the ansatz \eqref{eq:mbansatz}, the conserved charge \eqref{eq:saddle1} can be written as 
\begin{align}\label{eq:pcone}
    (\d_{\tau}f)^2 
    =  \frac{\Omega^2_k}{\lambda^2}\left( \lambda^2 f^2+1 \right)\left( \lambda^2 f^2+1 - \xi^2_k \right)
\end{align}
with the solution 
\begin{align}\label{eq:pboinsoln}
    f(\tau-\tau_0) = \lambda^{-1}\cs(\Omega_k(\tau-\tau_0),\xi_k)
    \,
\end{align}
implying the expression \eqref{eq:mbphi}. The parameters are given in \eqref{eq:mbpara1} and \eqref{eq:mbpara2}.

Due to the nature of $\cs(\Omega_k\tau,\xi_k^2)$, $f(\tau)$ is periodic with the period 
\begin{align}\label{eq:period1}
    \beta 
    =  \oint \frac{\Omega_k^{-1}\dd{h}}{\sqrt{(h^2+1)(h^2+1-\xi_k^2)}}
    = \Omega_k^{-1}\left[ 2pK(\xi_k^2) + 4qiK\left( 1-\xi_k^2 \right) \right]
\end{align}
in which we denote $f(\tau)$ for simplicity. 
Similar to the $\cp^1$ case, there exist four branch points $\pm i$ and $\pm i \sqrt{1-\xi^2_k}$ from which two branch cuts can be formed by linking $(-)i$ to $(-)i \sqrt{1-\xi^2_k}$, respectively. Consequently, the target space (of integration) is a torus characterized by a pair of integers $(p,q)$ representing the winding number around two cycles, $C_A$ and $C_B$, where $C_A$ spans the real line while $C_B$ is a cycle winding around $\pm i \sqrt{1-\xi^2_k}$.

Before diving into the calculation of the multibion solution's actions, let us take a closer look at the behavior of the system in the limit $\beta \to \infty$ with $(p,q) = (1,0)$. That is, taking $\xi_k^2 \to 1$, one arrives 
\begin{align}\label{eq:largebpara}
    \lambda^2 = \frac{\omega_k^2-m^2}{k\omega_k^2}
\end{align}
showing that 
\begin{align}
    \Omega_k = \omega_k 
    \qand
    E = \frac{\omega_k^2-m^2}{2g^2k} = m\epsilon \,.
\end{align}
This can also be deduced from the observation that as $\xi_k^2 \to 1$, $\cs(z,\xi_k^2)$ approaches $\csch{z}$, fulfilling the boundary condition specified in \eqref{eq:infbc}. This differs from the real bion solution only by a phase, which can be absorbed into the definition of $\alpha$.

With the general solution $f(\tau)$, we can then find the corresponding action and how it connects to the $p$-bion configurations in the large $\beta$ limit. To proceed, we plug \eqref{eq:pboinsoln} into the action \eqref{eq:actionqm}, leading to the expression
\begin{align}\label{eq:mbactiong}
    S_{\rm bion}(\beta,k) 
    &= -m\epsilon\beta + \frac{\Omega_k}{g^{2}}\oint \frac{\dd{h}}{\sqrt{(1+h^2)(1+h^2-\xi_k^2)}} \left[ \frac{\xi_k^2-1}{\lambda^2}  
        +\frac{2\lambda^2(1+h^2)(1+h^2-\xi_k^2)}{\lambda^4+2h^2\lambda^2k+h^4}
    \right]
\end{align}
where the latter integral (without specifying the corresponding cycles) results in
\begin{align*}
   \tilde{S}_1 =&
   \begin{multlined}[t][.95\linewidth]
        \frac{\Omega_k}{g^2}\;\Biggl\{
            \frac{1-\xi_k^2}{\lambda^2}F\left( \arcsin\sqrt{\frac{1}{h^2+1}},\xi_k^{2} \right)
        \\[2mm] 
        +\biggl[
            \frac{\lambda^2(k-\sqrt{k^2-1})-1+\xi_k^2}{\sqrt{k^2-1}}
            \Pi\left(1-\lambda^2(k-\sqrt{k^2-1}) ,\arcsin\sqrt{\frac{1}{h^2+1}},\xi_k^{2} \right)
        \\[2mm] 
        - \left( \sqrt{k^2-1} \to -\sqrt{k^2-1} \right)
        \biggr] 
        \Biggr\}
   \end{multlined} 
   \\
   =&
   \begin{multlined}[t][.95\linewidth]
        \frac{-i\Omega_k}{g^2\lambda^2}\;\Biggl\{
            (2\lambda^4-\xi_k^{\prime 2})F\left( \arcsin\sqrt{\frac{-h^2}{\xi_k^{\prime 2}}},\xi_k^{\prime 2} \right)
        \\[2mm] 
        +\Biggl[
            \left( \frac{\lambda^4(k-\sqrt{k^2-1})-\lambda^2(1+\xi_k^{\prime 2})+(k+\sqrt{k^2-1})\xi_k^{\prime 2}}{\sqrt{k^2-1}} \right) \times
        \\[2mm]
            \times\Pi\left( \frac{\xi_k^{\prime 2}}{\lambda^2(k-\sqrt{k^2-1})},\arcsin\sqrt{\frac{-h^2}{\xi_k^{\prime 2}}},\xi_k^{\prime 2} \right)
        - \left( \sqrt{k^2-1} \to -\sqrt{k^2-1} \right)
        \Biggr] 
        \Biggr\}
   \end{multlined} 
\end{align*}
In the last expression $\xi_k^{\prime 2}$ is the dual elliptic modulus $1- \xi_k^2$. Note that for two expressions of $\tilde{S}_1$, the imaginary-argument transformation \cite{NIST:DLMF,Byrd_1971} is applied such that the analytic properties is properly defined over the cycle $A$ and $B$, respectively. Note also that four poles emerge at 
\begin{align}
    h_{\pm\pm} = \pm \sqrt{\lambda^2 (-k \pm \sqrt{k^2-1}) }
\end{align}
in \eqref{eq:mbactiong}. With all ingredients at hand, we can express the bion action as 
\begin{align}\label{eq:bionaction1}
    S_{\rm bion}(\beta,k) = 
    -m\epsilon\beta 
    + pS_A + qS_B + 2\pi i lS_{\rm res}
\end{align}
where the subscript $A/B$ denotes the $C_{A/B}$ integration contour, $ S_{\rm res}$ is the residue at\footnote{The integration contour is defined on a torus with four punctures. Notice that each of two punctures collapses in the limit $k \to 1$.} $h_{\pm\pm}$.
The explicit form of $S_{A,B,{\rm res}}$ is presented as follows.
\begin{align*}
    S_A &= \begin{multlined}[t][.9\textwidth]
        \frac{-2\Omega_k}{g^2\sqrt{k^2-1}}\;\Biggl\{
            \frac{(1-\xi_k^2)\sqrt{k^2-1}}{\lambda^2}K(\xi_k^{2})
        \\
        +[\lambda^2(k-\sqrt{k^2-1})-1+\xi_k^2]\cdot
        \Pi\left(1-\lambda^2(k-\sqrt{k^2-1}) ,\xi_k^{2} \right)
        \\
        - [\lambda^2(k+\sqrt{k^2-1})-1+\xi_k^2]\cdot
        \Pi\left(1-\lambda^2(k+\sqrt{k^2-1}) ,\xi_k^{2} \right)
        \Biggr\}
    \end{multlined} 
   \\
    S_B &= \begin{multlined}[t][.9\linewidth]
        \frac{4i\Omega_k}{g^2\lambda^2\sqrt{k^2-1}}\;\Biggl\{
            \sqrt{k^2-1}(2\lambda^4-\xi_k^{\prime 2})K(\xi_k^{\prime 2})
        \\
        + [\lambda^4(k-\sqrt{k^2-1})-\lambda^2(1+\xi_k^{\prime 2})+(k+\sqrt{k^2-1})\xi_k^{\prime 2}] \cdot
        \Pi\left( \frac{\xi_k^{\prime 2}}{\lambda^2(k-\sqrt{k^2-1})},\xi_k^{\prime 2} \right)
        \\ 
        - [\lambda^4(k+\sqrt{k^2-1})-\lambda^2(1+\xi_k^{\prime 2})+(k-\sqrt{k^2-1})\xi_k^{\prime 2}] \cdot
        \Pi\left( \frac{\xi_k^{\prime 2}}{\lambda^2(k+\sqrt{k^2-1})},\xi_k^{\prime 2} \right)
        \Biggr\}
    \end{multlined} 
    \\[2mm]
    S_{\rm res} &= 
    \begin{multlined}[t][.9\linewidth]
        \frac{\pm \Omega_k}{2\lambda g^2\sqrt{k^2-1}} \Biggl[
            \sqrt{-\lambda^4(k+\sqrt{k^2-1})+\lambda^2(2-\xi_k^2)+(\xi_k^2-1)(k-\sqrt{k^2-1})}
            \\
            - 
            \sqrt{-\lambda^4(k-\sqrt{k^2-1})+\lambda^2(2-\xi_k^2)+(\xi_k^2-1)(k+\sqrt{k^2-1})}
        \Biggr]
        \end{multlined}
\end{align*}

Back to the large $\beta$ limit, the action turns out to be 
\begin{align}\label{eq:lbaction}
    S_{\rm bion}(\beta\to\infty,k)
    = p \srb
    + 2\pi il S_{{\rm res},\infty}
\end{align}
where the vacuum value $-m\epsilon\beta$ is neglected and the expression of $\lambda$ in \eqref{eq:largebpara} is substituted.
In such a limit\footnote{
    The limit of elliptic $\Pi$-function is taken by utilizing Eqn. \eqref{eq:jpiid}.
}, $S_B$ vanishes and $2\pi i \cdot S_{{\rm res},\infty}$ takes the same form as \eqref{eq:cbrbdiff}.

\section{Energy corrections from path integral formulation}
\label{sec:apppi}

In this appendix, we present detailed calculations for deriving the ground state energy from both single and multi-bion saddle configurations.

Let us start with the flow equation \eqref{eq:floweq}. By rescaling the ``time'' 
\begin{align}
    t \to \frac{\log{R_0}}{\sqrt{k^2-1}}t
    \,,
\end{align}
one arrives at the flow equation of the $\cp^1$ model \cite{Fujimori:2016ljw}. This is consistent with the fact that the effective action \eqref{eq:Seffkk} has the same dependence on the coordinates in the quasi-moduli space as the $\cp^1$ model. Here we quote the result below for completeness.
The thimble solutions are 
\begin{align}
\tau(t) = \frac{1}{m}\log[\frac{2m}{\epsilon g^2}(1+e^{\epsilon m (t-t_0)})]
+ \frac{i}{m}(\sigma\pi-\theta)
\end{align}
for $\sigma\in\ZZ$. The coefficient $n_{\sigma}(\theta)$ in $\cC = \sum_{\sigma}n_{\sigma}(\theta)\cJ_{\sigma}(\theta)$ can be found by the intersection between the original contour with the dual thimble $\cK_{\sigma}$

\subsection*{Multibion background}

We now outline the procedure for calculating the multi-bion corrections in the deformed model originating from the $\cp^1$ model. Importantly, the effective action \eqref{eq:Seffkk} for a kink-anti-kink pair retains the quasi-moduli structure characteristic of the standard $\cp^1$ model. This structure extends to the effective action for configurations with $p$ kinks and $p$ anti-kinks, allowing us to leverage similar methods in evaluating their contributions. To adapt this framework for the deformed model, it is sufficient to adjust the mass term in the effective potential, replacing the factor $m/g^2$ to $\ms/g^2$ in the $p$-bion configuration. As a result, the ratio $Z_p/Z_0$ \cite{Fujimori:2017oab} can be expressed as 
\begin{multline}\label{eq:pbionZ}
\left( \frac{Z_p}{Z_0} \right)_{\pm} = 
-\frac{im\beta}{p}e^{-\srbz} 
\\
\times\lim_{\sigma \to 0} \left( \pdv{\sigma} \right)^{p-1} \left[  
    \frac{8im^2}{g^4}e^{\frac{-i\sigma m \beta}{p}}
    \left( \frac{2m}{g^2}e^{\pm\frac{\pi i}{2}} \right)^{2(i\sigma-\epsilon)}
    \frac{\Gamma(\epsilon-\frac{i\sigma}{2})\Gamma(1-\frac{i\sigma}{2})}{\Gamma(1-\epsilon+\frac{i\sigma}{2})\Gamma(1+\frac{i\sigma}{2})}
\right]^p
\end{multline}
capturing the multi-bion effects within the modified potential landscape.
Recall that in the $\cp^1$ model we have the effective action $2m/g^2$ while in the deformed case it is $\srbz=2m_{*}/g^2$.

The leading order term of \eqref{eq:pbionZ} in the large $\beta$ limit indicates that the ground state energy takes the form 
\begin{align}
\tilde{E}_2 = 
-\lim_{\beta \to \infty} \frac{1}{\beta} \log\Biggl[
    \exp(\frac{2m\beta\Gamma(\epsilon)}{\Gamma(1-\epsilon)}e^{-\srbz\mp\pi i \epsilon}\left( \frac{2m}{g^2} \right)^{2(1-\epsilon)})
    + \cdots
\Biggr]
\end{align}
This indeed sketches the characteristics of the possible multibion expansion.
Lastly, the exact result around the nearly supersymmetric regime can be obtained from \eqref{eq:pbionZ}, following \cite{Fujimori:2017oab} 
\begin{align}
\label{eq:pbion1o}
\lim_{\epsilon\to 1}\tilde{E}_{p}^{(1)} &= -2m
\\
\label{eq:pbion2o}
\lim_{\epsilon\to 1}\tilde{E}_{p}^{(2)} &= 4mp^2\left( \gamma + \log\frac{2m}{g^2} \pm \frac{i\pi}{2} \right)
\end{align}

\section{Detail of \boldmath{$\delta\epsilon$} perturbation}
\label{sec:pertb1}

In this part, we detail the calculations of the perturbation of the ground state energy around the supersymmetric point. In consideration of the chosen moment map $\mu$, the normalization of the wave function is 
\begin{align}
    \braket{0}{0} &\equiv 
    \int\dd{\rho}\abs{\Psi^{(0)}}^2
    = \frac{\pi g^2}{2m}\left( 1-e^{-\srbz} \right)
\end{align}
in which the integral is evaluated over the measure 
\begin{align*}
    \dd{\rho} = \frac{i}{2}\frac{\dd{\varphi}\dd{\bphi{}}}{1+2k\abs{\varphi}^2+\abs{\varphi}^4}
    \,.
\end{align*}
Then, with the ansatz $\Psi^{(1)} = e^{-\mu/g^2}I(\abs{\varphi}^2)$, Eq. \eqref{eq:degfnone} becomes 
\begin{align}\label{eq:odewfa}
    XI^{\prime\prime} + \left( 1-\frac{2X\mu'}{g^2} \right)I'
    + g^{-2}\left( GE^{(1)} + \mu'+X\mu^{\prime\prime} \right)
    = 0
\end{align}
Note that $I$ is a function of $X\equiv \abs{\varphi}^2$ and $f'(X) \equiv \dv*{f}{X}$. One can solve the first-order differential equation \eqref{eq:odewfa} by the method of integration factor and have 
\begin{multline}\label{eq:D3}
    I'(X) 
    = -\frac{e^{2\mu/g^2}}{g^2X}   \Biggl[
            \frac{g^2\sqrt{k^2-1}R_0E^{(1)}}{m(1-R_0^2)}e^{-2\mu/g^2} 
            \\[2mm]
    - \frac{mg^2\sqrt{k^2-1}R_0e^{-2\mu/g^2}}{(1-R_0^2)^2} \left(  
                \frac{\exp(\frac{2\mu\sqrt{k^2-1}}{m})}{g^2\sqrt{k^2-1}-m}
                + \frac{\exp(-\frac{2\mu\sqrt{k^2-1}}{m})}{g^2\sqrt{k^2-1}+m}R_0^2
            \right)
            +C
        \Biggr]
\end{multline}
where the constant $C$ is 
\begin{align}
    C = \frac{m^2}{g^4(k^2-1)-m^2}\frac{2m^2g^2R_0\sqrt{k^2-1}}{1-R_0^2}
    \left( 1 - e^{\srbz} \right)^{-1}
\end{align}
to make $I'(X)$ regular at $X=0$. This therefore gives the first order modification of the wave function.

\subsection*{Second order calculations}

For the task of computing the second order correction to the energy, it is convenient to recast \eqref{eq:rs2nde} and \eqref{eq:D3} in the form 
\begin{align}\label{eq:E2gen}
    E^{(2)} = \frac{1}{\braket{0}{0}}\mel**{\Psi^{(1)}}{\delta{H}-E^{(1)}}{0}
\end{align}
and 
\begin{align}\label{eq:wf1corr}
    \braket{\Psi^{(1)}}{0} = 
    \frac{\pi g^2R_0\sqrt{k^2-1}}{m(1-R_0^2)}
    \left[  
    \tilde{I}(\ms)
    -
    \int_{0}^{\ms}\dd{\mu}
    e^{-\frac{2\mu}{g^2}}\tilde{I}'
    \right]
\end{align}
where $\tilde{I}(\mu) = I(R)(\dv*{R}{\mu})$ and $\ms = m\log R_0/\sqrt{k^2-1}$.
After substituting $\delta{H}$, \eqref{eq:wf1corr}, and \eqref{eq:eeonepnp} in \eqref{eq:E2gen}, one obtains the first line of \eqref{eq:ambstr}, say,
\begin{align}\label{eq:e2ambfull}
\frac{\cosh{\frac{\ms}{g^2}}}{\sinh^3{\frac{\ms}{g^2}}} \Biggl[
\gamma + \log\frac{2\ms}{g^2}
\pm \frac{i\pi}{2}
+ \frac{1}{2}\int_{0}^{\infty} e^{-t}
\left( 
    \frac{e^{2\ms/g^2}}{t-\frac{2\ms}{g^2 \pm i0}}
+ \frac{e^{-2\ms/g^2}}{t+\frac{2\ms}{g^2}}
 \right)
 \dd{t}
\Biggr]
\end{align}
where $\gamma$ is the Euler-Mascheroni constant. The third and the fourth term correspond to the ambiguity. Together with the overall coefficient, one arrives Eqn. \eqref{eq:ambE}.

Next, let us confirm that any ambiguity originates solely from the first line of Eqn. \eqref{eq:ambstr}. After completing the integration, we can explicitly express the second line of Eqn. \eqref{eq:ambstr} $\tilde{I}_1$ as 
\begin{align}\label{eq:nambck}
\tilde{I}_{1} &= \frac{m}{\sqrt{k^2-1}} \sum_{n=1}^{\infty} \frac{2^{n}B_{n}(2)}{\Gamma(n+1)}\left( \frac{\sqrt{k^2-1}}{m} \right)^{n}\left[ 1 + (-1)^{n}e^{-\srbz} \right]
\cdot \frac{\ms^{n}\left[ \oFt{\frac{n}{2};\frac{n+2}{2},\frac{1}{2};\frac{\ms^2}{g^4}}-1 \right]}{n}
\end{align}
where $\oFt{a;b,c;z}$ is the generalized hypergeometric function. 
Since the latter term
\begin{align}
    \frac{\ms^{n+1}\oFt{\frac{n+1}{2};\frac{n+3}{2},\frac{1}{2};\frac{\ms^2}{g^4}}}{n+1}
    \sim 
    \left( \frac{g^2}{2} \right)^{n+1}
    \times (\cdots)
\end{align}
implying $\tilde{I}_1$ is a series in $g^2$, we have to take a closer look to ensure no ambiguity shows up after resummation. Through the explicit expression of the hypergeometric function $\oFt{a;b,c;z}$, \eqref{eq:nambck} turns out to be 
\begin{multline}\label{eq:Iamb1}
\tilde{I}_1 = \frac{m}{\sqrt{k^2-1}} \Biggl\{
    - \sum_{n=1}^{\infty}\tilde{A}(n) \cdot \frac{\ms^{n}}{n}
    +\sum_{n=1}^{\infty} \left( \frac{g^2}{2} \right)^{2n}\tilde{A}(2n)\Gamma(2n)
    \\[2mm]
    + \frac{g^2}{2}\sinh(\frac{2\ms}{g^2})
    \cdot \frac{1}{2}\left[ F(\ms,g^2) + F(\ms,-g^2) \right]
    \\[2mm]
    - \frac{g^2}{2}\cosh(\frac{2\ms}{g^2})\cdot \frac{1}{2}\left[ F(\ms,g^2) - F(\ms,-g^2) \right]
\Biggr\}
\end{multline}
where 
\begin{align*}
\tilde{A}(n) &\equiv
\frac{2^{n}B_{n}(2)}{\Gamma(n+1)}\left( \frac{\sqrt{k^2-1}}{m} \right)^{n}\left[ 1 + (-1)^{n}e^{-\srbz} \right] 
\end{align*}
and 
\begin{align*}
    f(\ms) \equiv \sum_{n=1}^{\infty} \tilde{A}(n)\ms^{n-1}
    \qand
    F(\ms,g^2) \equiv \sum_{s=0}^{\infty}\left( \frac{g^2}{2}\dv{\ms} \right)^s f(\ms)
    \,.
\end{align*}
Now, \eqref{eq:Iamb1} is simple enough for further considerations. First, let us analyze the asymptotic behaviors of $\tilde{A}(n)$, and $F(\ms,g^2)$. To proceed, by Eqn. \eqref{eq:berasym}, we know that 
\begin{align}
    \tilde{A}(n) \sim \frac{1}{\Gamma(n+1)}\left( \frac{2\sqrt{k^2-1}}{m} \right)^{n}\left[ 1 + (-1)^{n}e^{-\srbz} \right]
    \cdot 
    \frac{2\Gamma(n+1)}{(2\pi)^n}
\end{align}
when $n$ is large. It is not factorial since $\Gamma(n+1)$ cancels with each other. From this we also know that $f(\ms)$ should not be factorial as well because it is nothing but a $\ms$ series with the coefficient $\tilde{A}(n)$.
As for $F(\ms,g^2)$, the additional component comes from $(\dv*{\ms})^{j}(\ms)^n$. This generates the factor 
\begin{align}
    (n)_{j} = n(n+1)(n+2)\cdots(n+j-1)
    \sim n^j + \frac{1}{2}n^{j-1}j(j-1)
    \qas
    n\to\infty
\end{align}
which is again not factorial. We then conclude that the last two lines in \eqref{eq:Iamb1} is not factorial and hence produces no ambiguity. Only the first line in \eqref{eq:Iamb1} has to be treated carefully, especially the second term.

For the first term, since there is no $g^2$ dependence, it does not have any ambiguity in $g^2$ perturbative series. For the second term, it can be recast in the form 
\begin{align}\label{eq:namb2nd}
\sum_{n=1}^{\infty} \left( \frac{g^2}{2} \right)^{2n}\tilde{A}(2n)\Gamma(2n)
= 
\frac{1}{2}\left[  
    \sum_{n=1}^{\infty} \left( \frac{g^2}{2} \right)^{n}\tilde{A}(n)\Gamma(n)
    + 
    \sum_{n=1}^{\infty} \left( -\frac{g^2}{2} \right)^{n}\tilde{A}(n)\Gamma(n)
\right]
\end{align}
and the associated Borel representation is 
\begin{multline}\label{eq:I2}
\frac{2}{g^2}\int_{0}^{\infty} \dd{t} \, e^{-2t/g^2}\cB{I_2}[t]
= 
\frac{2}{g^2}\int_{0}^{\infty} \dd{t} \, e^{-2t/g^2}(1+e^{-\srbz})
\\[2mm]
\times\Biggl[
    \frac{te^{-2t\sqrt{k^2-1}/m}\sqrt{k^2-1}}{m}\frac{e^{6t\sqrt{k^2-1}/m}+1}{e^{-2t\sqrt{k^2-1}/m}-1}
    -1
\Biggr]
\end{multline}
where $\cB{I_2}(t)$ denotes the Borel transformation of Eqn. \eqref{eq:namb2nd}. One can immediately see that this term is Borel-summable and has \emph{no} ambiguity since it has no pole on the non-negative real line.\footnote{Note that \eqref{eq:I2} converges because it is finite at $t=0$ and as $t \to \infty$
\begin{align*}
    \dd{I_2} \sim t \exp[-2t\left( \frac{1}{g^2} - \frac{\sqrt{k^2-1}}{m} \right)]
    \to 0
\end{align*}
which in our previous discussion we have $m > 2g^2\sqrt{k^2-1}$.}

Lastly, it is also noteworthy to provide the Borel representation of $F(\ms,\pm g^2)$. That is, 
\begin{multline}
F(\ms,g^2) = \frac{2}{g^2} \int_{0}^{\infty}\dd{t}\, 
\frac{e^{-2t/g^2}}{\pm t+\ms}\Bigl[ 
    \GB{\pm t+\ms,2} 
    \\
    + e^{-\srbz}\GB{\pm t+\ms,-1}
    - 
    (1+e^{-\srbz})
\Bigr]
\end{multline}
where $G_B(t,x)$ is the generating function of the Bernoulli polynomials defined in Appendix \ref{sec:nc}. Note that one may think that the ambiguity arises at $t=\ms \in [0,\infty)$. However, this is not the case since the residue at $t=\ms$ is zero.

Summarizing, we showed that the ambiguity essentially arises from the $\mu^{-1}$ term in the expansion, leading to \eqref{eq:e2ambfull}. 
The higher order terms associated to the Lie-algebraic parameter $k^2-1$ are Borel summable and produce no ambiguity.
This indicates that the ambiguity in the Lie-algebraic deformed model stems from the same root as that of the $\cp^1$ model.

\newpage
\section*{}\addcontentsline{toc}{section}{References}
{\providecommand{\href}[2]{#2}
\providecommand{\doihref}[2]{\href{#1}{#2}}
\providecommand{\arxivfont}{\rm}

}

\end{document}